%% file: main.tex
\documentclass[review]{elsarticle}

\usepackage{lineno,hyperref}
\usepackage{amsthm,amsmath}
\usepackage{amsfonts}
\usepackage{times}
\usepackage{graphicx}
\usepackage{color}{\tiny }
\usepackage{multirow}
\usepackage{hyperref}
\usepackage{color,soul}
\usepackage{xr}
\usepackage[utf8]{inputenc}
\usepackage{rotating}
\usepackage{bbm}
\usepackage{latexsym}
\usepackage{braket}

\usepackage[justification=justified,singlelinecheck=false,font=small,labelfont=bf]{caption}
\usepackage{subcaption}
\usepackage{import}

\newcommand{\N}{N} 

\newcommand{\E}{\mathbb{E}}
\newcommand{\expect}[1]{\langle #1 \rangle}
\newcommand{\ev}[1]{\langle #1 \rangle}

\newcommand{\R}{\mathbb{R}}
\newcommand{\cov}[0]{\text{cov}}

\renewcommand{\P}{\mathbb{P}}
\newcommand{\Q}{\mathbb{Q}}
\newcommand{\Pmeasure}{\mathbb{P}}
\newcommand{\Qmeasure}{\mathbb{Q}}

\newcommand{\X}[0]{X} 
\newcommand{\Y}[0]{Y} 

\newcommand{\norm}[1]{\left\Vert #1 \right\Vert}
\newcommand{\tnm}{{t_{n-1}}}

\journal{Journal of Mathematical Psychology}

\biboptions{authoryear}
\begin{document}

\begin{frontmatter}

\title{The Hitchhiker's Guide to Nonlinear Filtering}


\author[mynewaddress,mymainaddress]{Anna Kutschireiter\corref{mycorrespondingauthor}}
\cortext[mycorrespondingauthor]{Corresponding author, anna.kutschireiter@gmail.com}

\author[mymainaddress,mysecondaryaddress]{Simone Carlo Surace}

\author[mymainaddress,mysecondaryaddress]{Jean-Pascal Pfister}

\address[mynewaddress]{Department of Neurobiology, Harvard Medical School, 200 Longwood Avenue, Boston, MA 02115, United States.}
\address[mymainaddress]{Department of Physiology, University of Bern, B\"{u}hlplatz 5, 3012 Bern, Switzerland.}
\address[mysecondaryaddress]{Institute of Neuroinformatics, UZH / ETH Zurich, Winterthurerstrasse 190, 8057 Zurich, Switzerland.}

\begin{abstract}
Nonlinear filtering is the problem of online estimation of a dynamic hidden variable from incoming data and has vast applications in different fields, ranging from engineering, machine learning, economic science and natural sciences.
We start our review of the theory on nonlinear filtering from the simplest `filtering' task we can think of, namely static Bayesian inference.
From there we continue our journey through discrete-time models, which is usually encountered in machine learning, and generalize to and further emphasize continuous-time filtering theory.
The idea of changing the probability measure connects and elucidates several aspects of the theory, such as the parallels between the discrete- and continuous-time problems and between different observation models.
Furthermore, it gives insight into the construction of particle filtering algorithms.
This tutorial is targeted at scientists and engineers and should serve as an introduction to the main ideas of nonlinear filtering, and as a segway to more advanced and specialized literature.
\end{abstract}


\end{frontmatter}


\input{01_Introduction.tex}

\input{02_A-view-from-space.tex}

\input{03_Filtering-formal.tex}

\input{04_Approximate-closed-form.tex}

\input{05_Particle-filters.tex}

\input{08_Conclusion.tex}

\section*{So long, and thanks for all the fish: Acknowledgments}

The authors would like to express their gratitude towards Christian Horvat, Jannes Jegminat, Luke Rast, Hazem Toutounji, and the anonymous reviewers for taking a test ride and providing constructive feedback.
Further, we thank Johannes Bill and Jan Drugowitsch for helpful discussions.

\section{Funding}

AK, SCS and JPP  were supported by the Swiss National Science Foundation grants PP00P3\_179060 and PP00P3\_150637 (AK, SCS, JPP) and P2ZHP2\_184213 (AK).

 \bibliographystyle{apalike}
\bibliography{library}

\input{07_Appendix.tex}

\end{document}

%% file: 01_Introduction.tex
\section{Introduction: A Guide to the Guide}

\begin{flushright}
	\textit{``The introduction begins like this:\\Space, it says, is big. Really big. \\You just won’t believe how vastly hugely mind-bogglingly big it is.''}\\
	--- Douglas Adams
\end{flushright}

Filtering is the problem of estimating a dynamically changing state, which cannot be directly observed, from a stream of noisy incoming data.
To give a concrete example, assume that you are a Vogon in charge of a spaceship.
Since you had a particularly bad day, you decide to destroy a small asteroid to make you feel better.
Before you push the red button, you need to know the current position of the asteroid, which corresponds to the hidden state $X_t$.
You have some idea about the physics of movement in space, but there is also a stochastic component in the movement of your target.
Overall, the asteroid's movement is described by a stochastic dynamical model.
In addition, you cannot directly observe its position (because you like to keep your safe distance), so you have to rely on your own ship's noisy measurements $Y_t$ of the position of the asteroid.
Because of these uncertainties, it would not only be useful to have an estimate of the asteroid's current position $X_t$ based on the history of measurements $ \Y_{0:t} = \{ Y_0, Y_1, \dots , Y_t \}$, but also an estimate of the uncertainty of the estimate.
Thus generally, the conditional probability density $ p(X_t | \Y_{0:t}) $ is the complete solution to your problem (and the beginning of the problem of how to find this solution).

These sorts of problems are not only relevant for bad-tempered Vogons, but in fact are encountered in a wide variety of applications from different fields.
Initial applications of filtering were centered mostly around engineering.
After the seminal contributions to \emph{linear} filtering problems by \citet{Kalman1960,Kalman1961}, the theory was largely applied to satellite orbit determination, submarine and aircraft navigation as well as space flight \citep{Jazwinski1970}.
Nowadays, applications of (nonlinear) filtering range from engineering, machine learning \citep{Bishop}, economic science (in particular mathematical finance, some examples are found in \citealp{Brigo1998a}) and natural sciences such as geoscience \citep{VanLeeuwen2010}, in particular data assimilation problems for weather forecasting, neuroscience and psychology.
As a particular example for its usefulness in neuroscience, the modeling of neuronal spike trains as point processes \citep{Brillinger1988a,Truccolo2005} has led to interesting filtering tasks, such as the problem of decoding a stimulus from the spiking activity of neurons (e.g.~\citealp{Koyama2010,Macke2011}).
In psychology, nonlinear filtering techniques are not only used for data analysis, but can also provide qualitative insight into psychological processes such as perception \citep{Wolpert1995,Kording2007} or decision making \citep{Drugowitsch2014a,Glaze2015,Radillo2017,Veliz-Cuba2016,Piet2017}.
To tackle these kinds of questions, knowledge about nonlinear filtering is indispensable.
Theoretical understanding can further help in connecting and unifying specific applications of filters and is paramount for understanding more advanced topics in filtering \citep[Section 1.2]{Jazwinski1970}.

The aim of this tutorial is to present -- in an easily accessible and intuitive way -- the theoretical basis for continuous-time nonlinear filtering with diffusion and point-process observations.
The tutorial highlights the change of measure as a powerful tool to derive the fundamental equations of nonlinear filtering as well as numerical approximations. 
In addition, the unification provided by the concept of change of measure provides a solid basis for diving into the huge body of literature on nonlinear filtering.
Our tutorial complements the more advanced theoretical literature (e.g.~\citealp{Jazwinski1970,Bain2009,bremaud1981point}) or more specialized tutorials, e.g.~on particle filtering \citep{Arulampalam2002,Doucet2009,Speekenbrink2016}, Hidden Markov Models \citep{Visser2011,Rabiner1989} or variational Bayes for latent linear systems \citep{Ostwald2014}.

%% file: 02_A-view-from-space.tex
\section{A view from space: from Bayes' rule to filtering}
\label{sec:VfS}
\begin{flushright}
\textit{``Even the most seasoned star tramp can’t
help but shiver\\at the spectacular drama of a sunrise seen from space,\\
but a binary sunrise is one of the marvels of the Galaxy.''}\\
--- Douglas Adams
\end{flushright}
Suppose that we observe a random variable $Y$ and want to infer the value of an (unobserved) random variable $X$.
Bayes' rule tells us that the conditional distribution of $X$ given $Y$, the so-called posterior, can be computed in terms of three ingredients:
the prior distribution $p(X)$, the likelihood $p(Y|X)$, and the marginal likelihood $P(Y)$ (which acts as a normalizing constant):
\begin{eqnarray}
p(X|Y)=\frac{p(Y|X)p(X)}{p(Y)}.
\label{eq:Intro Filt - Bayes' rule}
\end{eqnarray}
This tutorial is concerned with the application of the above idea to a situation where $X,Y$ are continuous-time stochastic processes and we want to perform the inference online as new data from $Y$ comes in. 
In this section, we want to gradually build up the stage: 
as some readers might be more familiar with discrete-time filtering due to its high practical relevance and prevalence, we will start our journey from there, picking up important recurring concepts as we make our way to continuous-time filtering.

\subsection{Changes of measure}
\label{sec:Changes of measure}

Before we talk about dynamic models, let us briefly highlight a concept in Bayesian inference that will be very important in the sequel: that of changing the probability measure.
A probability measure is a function that assigns numbers (`probabilities') to events.
If we have two such measures $ \P $ and $ \Q $, then $\P$ is called \emph{absolutely continuous} wrt. $\Q$ if every nullset of $\Q$ is a nullset of $\P$.
Moreover, $\P$ and $\Q$ are called \emph{equivalent} if they have the same nullsets.
In other words, if $ A $ denotes an event, and $ P(A) $ denotes its probability, then equivalence means that $ \Q(A) = 0 $ if and only if  $ \P(A) = 0 $.

But why would we want to change the measure in the first place?
Changing the measure allows us to compute expectations of a measurable function $ \phi(x) $ with respect to a measure $ \Q $, which were originally expressed with respect to another measure $ \P $.
To see this, consider the two measures $ \P $ and  $ \Q $ for some real-valued random variable $ X $, and write them in terms of their densities $p,q$ (with respect to the Lebesgue measure).\footnote{In this section, all `densities' are with respect to the Lebesgue measure.}
We then have
\begin{eqnarray}
	\E_\P \left[ \phi(X) \right] & = & \int  dx \,  \phi(x) p(x) \nonumber 
	\\
	& = & \int \, dx \,  \frac{p(x)}{q(x)}  \phi(x)  q(x)  = \E_\Qmeasure \left[ L(X)  \phi(X) \right],
	\label{HeuristicRN}
\end{eqnarray}
where we introduced the likelihood ratio $ L(x) := \frac{p(x)}{q(x)}  $ and $ \E_\Q $ denotes expectation under the distribution $ q $.
Thus, changing the measure proves to be very useful whenever expectations under $\Q$ are easier to compute than under $\P$.


A fundamental problem in filtering is that of computing a conditional expectation (i.e. an expected value under the posterior distribution) of this sort:
\begin{eqnarray}
\E_\P[\phi(X)|Y] &=& \int dx \, \phi(x) p(x|Y)
\label{eq:Intro Filt - Posterior exp}
\end{eqnarray}
for some function $\phi$, and we want to use Eq.~\eqref{eq:Intro Filt - Bayes' rule} to compute $p(X|Y)$.
We therefore have to compute the two integrals here
\begin{eqnarray}
\mathbb{E}_\P[\phi(X)|Y] & = & \int dx \, \phi(x) \frac{p(Y|x) p(x)}{p(Y)} = \frac{\int dx \, \phi(x) p(Y|x)p(x)}{\int dx \, p(Y|x)p(x)},
\label{eq:Intro Filt - Bayes' rule exp}
\end{eqnarray}
but the structure of the model (interactions between $X$ and $Y$) might make it very hard to compute the integrals, either analytically or numerically. 
Thus, we again change the measure to a reference measure $ \Q $ with joint density $ q(x,y) $, and rewrite Eq.~\eqref{eq:Intro Filt - Bayes' rule exp}:
\begin{eqnarray}
	\mathbb{E}_\P[\phi(X)|Y] & = &  \frac{\int dx \, \phi(x) \frac{p(x,Y)}{q(x,Y)} q(x,Y)  }{\int dx \, \frac{p(x,Y)}{q(x,Y)}  q(x,Y) } = \frac{ \E_\Q [ L(X,Y) \phi(X) |Y  ] }{\E_\Q [ L(X,Y)  |Y  ]}, \label{eq:Intro Filt - Bayes' rule exp change of measure}
\end{eqnarray}
where now the likelihood ratio $ L(x,y)  =  \frac{p(x,y)}{q(x,y)} $ is a function of both $ x $ and $ y $.

The hope is that we can pick a reference measure $\Q$ such that both $L(x,y)$ and $q(x,y)$ are simple enough to make Eq.~\eqref{eq:Intro Filt - Bayes' rule exp change of measure} more tractable than Eq.~\eqref{eq:Intro Filt - Posterior exp}.
For instance, some simplification might be achieved by switching from a model $p(x,y)$ of $\P$ in which $X$ and $Y$ are coupled, i.e.~statistically dependent, to a model $p(x)q(y)$ of $\Q$ where they are independent (while preserving the distribution of $X$), i.e.~under model $ \Q $ we find $q(x,y)=p(x)q(y)$. 
A potential added advantage of changing measure is when the distribution $ q(y) $ is computationally simple.
Then, the likelihood ratio $L(x,y)$ reads
\begin{eqnarray}
L(x,y)= \frac{p(x,y)}{q(x,y)} = \frac{p(y|x)p(x)}{p(x)q(y)} = \frac{p(y|x)}{q(y)},
\label{eq:Intro Filt - Lratio}
\end{eqnarray}
and conditional expectations under $ \Q $ can simply be taken with respect to the prior probability $ p(x) $.

Please take a moment to appreciate the value of this idea: the change of measure has allowed us to replace the expectation with respect to the posterior $ p(x|y) $ of $ \P $ (which might be hard to get) with an expectation with respect to the prior $ p(x) $ of $ \Q $ (which might be easy to compute).
This `trick' will become the central theme of this manuscript.

\subsubsection{Importance sampling}
\label{sec:Importance sampling}

\begin{figure}
    \centering
    \includegraphics[width=0.9\textwidth]{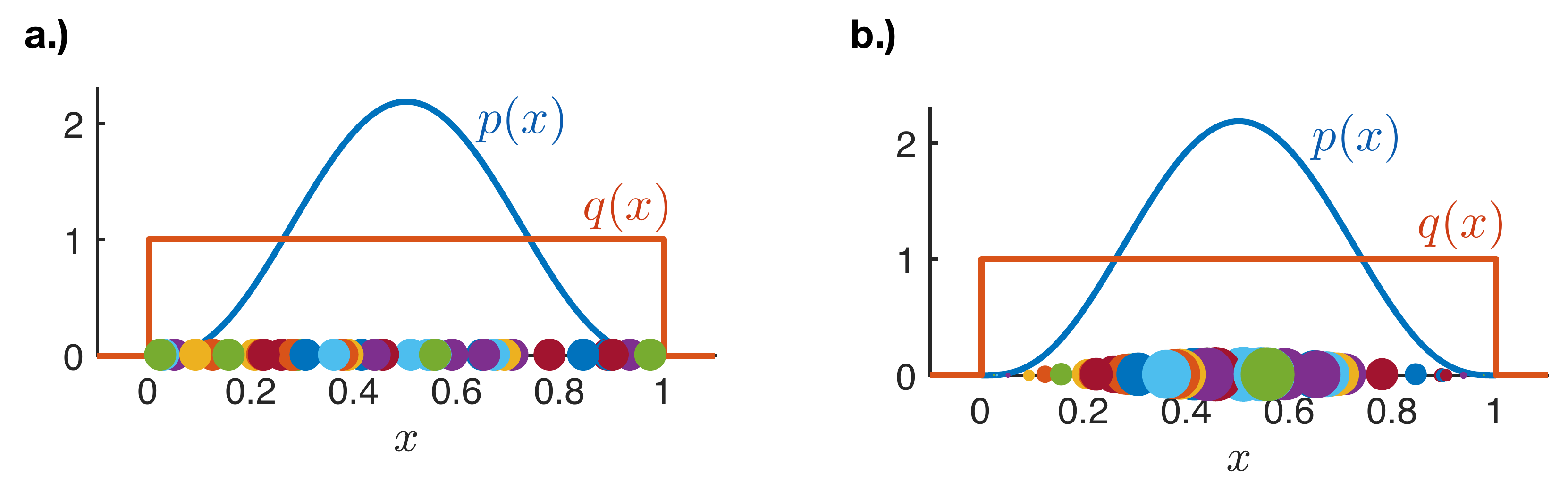}
    \caption{\label{fig:IS}
    Consider the problem of empirically approximating the beta distribution $p(x) = \text{Beta}(x;4,4)$ (blue) with samples from the uniform distribution between 0 and 1, $q(x)=\mathcal{U}(x;0,1)$ (red). 
    a.) The density of those samples does not represent the normal distribution, but b.) a combination of the density of samples together with their respective importance weights according to Eq.~\eqref{eq:importance sampling}.
    Here, the size of a dot represents the weight of the respective sample.}
\end{figure}

A numerical example where a `measure change'  is directly used is \textbf{importance sampling}.
Here, the goal is to approximate expectations with respect to a distribution $ p(x) $ (under $ \P $) by using $ M $ empirical samples $ X^i \sim p(x) $, such that
\begin{eqnarray}
	\E_\P [ \phi(X) ] & \approx & \frac{1}{M} \sum_{i=1}^M \phi(X^i).
\end{eqnarray}
However, there might be situations where we cannot draw samples from $ p(x) $, but only from another distribution $ q(x) $ (under $ \Q $).
Thus, we first perform a change of measure to $ \Q $, and then use the samples $ X^i \sim q(x) $ to approximate the expectation:
\begin{eqnarray}
	\E_\P [ \phi(X) ] & = & \E_\Q [ L(X) \,\phi(X) ] \approx \frac{1}{M} \sum_i L(X^i) \, \phi(X^i). \label{eq:importance sampling}
\end{eqnarray}
In this context, the likelihood ratio $ L(X^i) = \frac{p(X^i)}{q(X^i)} =: w_i $ is referred to as (unnormalized) \emph{importance weight}.
Hence, the target distribution $p(x)$ is not only represented by the \emph{density} of empirical samples (or `particles'), i.e.~how many samples can be found in a specific interval in the state space, but also by their respective importance weights (see simple example in Figure \ref{fig:IS}).

Similarly, we can use importance sampling to approximate a posterior expectation $ \E_\P [ \phi(X) |Y] $ with samples from the prior distribution $ p(x) $.
For this, consider changing to a measure $ \Q $ with density $ q(x,y) = p(x) q (y) $, such that with the likelihood ratio in Eq.~\eqref{eq:Intro Filt - Lratio} we find
\begin{eqnarray}
	\E_\P [ \phi(X) |Y] & = & \frac{ \E_\Q [ L(X,Y) \phi(X) |Y  ] }{\E_\Q [ L(X,Y)  |Y  ]} \nonumber
	\\
	 &\approx & \frac{1}{Z} \sum_i p(Y|X^i) \phi (X^i) = \frac{1}{Z} \sum_i w_i \phi (X^i), \,\,\, X^i \sim p(x) \label{eq:VfS - PF static}
\end{eqnarray}
where the unnormalized importance weights are given by $ w_i = p(Y|X^i) $, and the normalization $ Z $ is given by
\begin{eqnarray}
	Z = \sum_i p(Y|X^i) = \sum_i w_i.
\end{eqnarray}
Thus, in order to approximate a posterior with empirical samples from the prior, each sample $ X^i \sim p(x) $ has to be weighted according to how likely it is that this particular sample has generated the observed value of the random variable $ Y $ by evaluating the likelihood $ p(Y|X^i) $ for this sample.
Generalizing this to a dynamical inference setting will give rise to the bootstrap particle filter, and we will show in Section \ref{sec:Intro Filt - Particle methods as approximate solutions to the filtering problem} how changing the measure in empirical sampling for a dynamical system results in dynamical equations for the particles and weights.

\subsection{Filtering in discrete time - an introductory example}

The inference problems in the previous section were of purely static nature.\label{here}\footnote{If you ask yourself why we needed \pageref{here} pages to get to this point, please bear with us: the concept of changing the measure is very straightforward in a static setting, and might help to grasp the (seemingly) more complicated applications in a dynamic setting later on.}
However, this being the Hitchhiker's guide to nonlinear \emph{filtering}, let us now start to consider dynamical models for filtering.

Filtering means computing the conditional distribution of the hidden state $X_{t}$ at time $ t $ using the observations up to that time $ \Y_{0:t} = \{ Y_0, \dots, Y_t \} $.
There are two important ingredients to this problem: 
first, the \emph{signal model} describes the dynamics of the hidden state $X_{t}$.
In order to perform the inference recursively, the usual minimum assumption for the \emph{hidden, or latent, state process} $X_{t}$ with state space $ S $ is that it is a first-order Markov process, which, roughly speaking, means that the (probability of the) current state just depends on the last state, rather than on the whole history.
In discrete time,\footnote{In discrete time $t_n=n\Delta t$.} we can write more formally
\begin{eqnarray}
	p(X_{t_n}|\X_{t_{0:n-1}})  & = & p(X_{t_n}|X_{t_{n-1}}).
\end{eqnarray}
Thus, the dynamics of the whole process is captured by the \emph{transition probability} $ p(X_{t_n}|X_{t_{n-1}})$, which is assumed to be known.

Second, the \emph{observation model} describes the (stochastic) generation of the observation process $ Y_{t_n} $, and is captured by the \emph{emission probability} $ p(Y_{t_n}|X_{t_n})$, which is also assumed to be known.
Together, the transition and emission probability form a so-called state space model (SSM).\footnote{Somewhat oddly, the name `state space model' usually refers to a model with continuous state space, i.e.~$X_t \in \R^n$, which is distinct from models with finite state space such as the Hidden Markov Model below. In this tutorial, the state space can both be discrete or continuous, and if necessary, will be further clarified in the text.}
With these ingredients, the filtering problem in discrete time reduces to a simple application of Bayes' rule (Eq. \ref{eq:Intro Filt - Bayes' rule}) at each time step, which may be written recursively:
\begin{eqnarray}
	p(X_{t_n} | \Y_{0:t_n} ) & = & \frac{ p(Y_{t_n}|X_{t_n}) p(X_{t_n} | \Y_{0:t_{n-1}} ) }{  p(Y_{t_n} | \Y_{0:{t_{n-1}}} ) }
	\\
	& = & \frac{ p(Y_{t_n}|X_{t_n}) \int_S  d x_{t_{n-1}}  \, p(X_{t_n} | x_{t_{n-1}}) p(x_{t_{n-1}} | \Y_{0:t_{n-1}} ) }{  \int_S dx_{t_n} \, p(Y_{t_n} | x_{t_n}) \int_S dx_{t_{n-1}} \,  p(x_{t_n} | x_{t_{n-1}}) p(x_{t_{n-1}} | \Y_{0:t_{n-1}} ) }.\label{eq:VfS - filtering Bayes}
\end{eqnarray}

\begin{figure}
  \centering
      \includegraphics[width=0.9\textwidth]{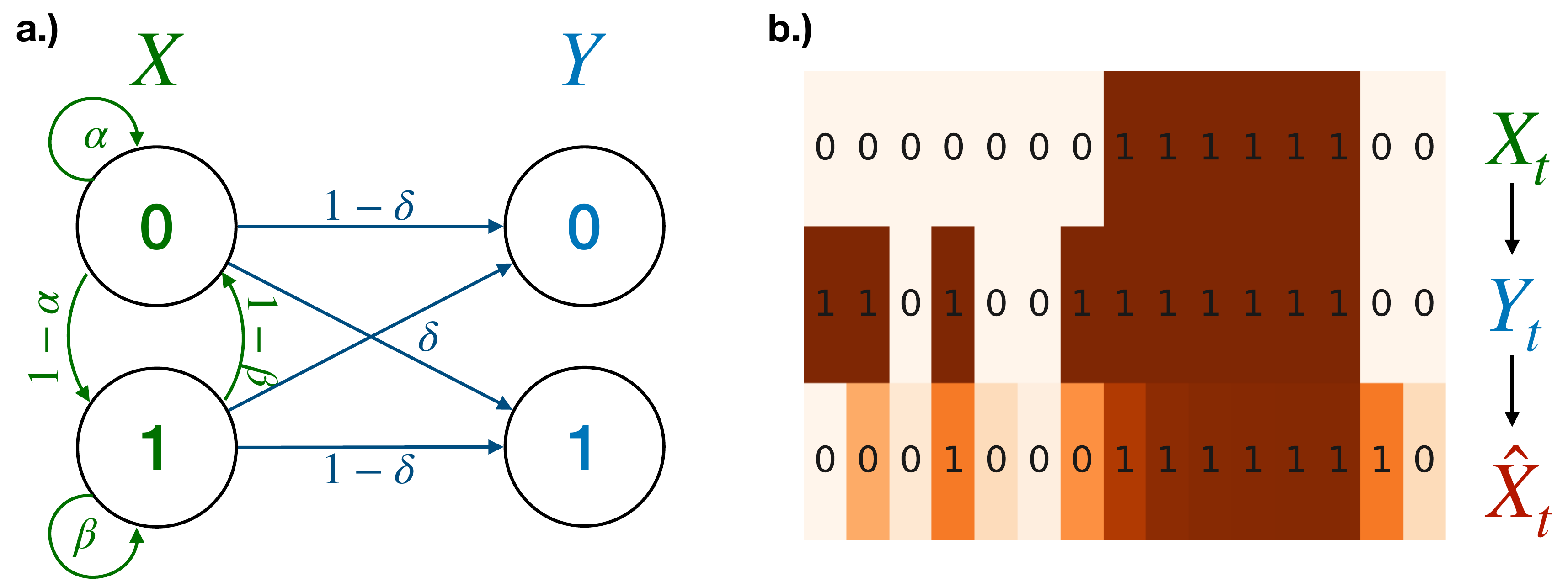}
  \caption{a.) A two-state HMM with binary observation channel. 
  $\alpha$ and $\beta$ denote the probability to stay in state $0$ and $1$, respectively.
  The probability of making an error in the observation channel is given by $\delta$.
  b.) Sample state trajectory, sample observation and filtered density $\hat{p}_{t_n}$ (color intensity codes for the probability to be in state $1$), as well as estimated state trajectory $ \hat{X}_{t_n} $ (where $ \hat{X}_{t_n} = 1 $ if $\hat{p}_{t_n }^{(2)} > 1/2 $ and $ \hat{X}_{t_n} = 0$ otherwise).
  \label{fig:HMM}}
\end{figure}

The simplest dynamic model for filtering is a Hidden Markov Model (HMM).
To see that you don't need rocket science for hitchhiking and applying Eq.~\eqref{eq:VfS - filtering Bayes}, let us consider an HMM with two hidden states and two observed states, i.e. $X_{t_n}$ and $Y_{t_n}$ can take values of 0 or 1 for each time $t_n$.
The transition probabilities for $X_{t_n}$ are given by 
\begin{eqnarray}
p(X_{t_{n}}=0|X_{t_{n-1}}=0)=\alpha, \quad p(X_{t_{n}}=1|X_{t_{n-1}}=1)=\beta.
\end{eqnarray}
Thus, $\alpha$ is the probability of staying in state 0, whereas $\beta$ is the probability of staying in state $1$, and leaving those states has to have probability $1-\alpha$ and $1-\beta$, respectively, where we assume that $0<\alpha,\beta<1$ such that each state is visited.
This can be represented by a matrix.\footnote{Here $P^{\top}$ is used to denote the transpose of the matrix $P$.
It will become clear later why we define the transition matrix as $P^{\top}$.}
\begin{eqnarray}
P^{\top}=\begin{pmatrix}
\alpha & 1-\beta \\
1-\alpha & \beta
\end{pmatrix}, \label{eq:HMM transition}
\end{eqnarray}
which recursively determines the distribution of the hidden Markov chain at each time:
if $p_{t_{n-1}} = (p_{t_{n-1}}^{(1)},p_{t_{n-1}}^{(2)})^\top$ is a two-dimensional vector, denoting probability of state occupancy at time $t-1$, i.e.~$ p_{t_{n-1}}^{(1)} = P(X_{t_{n-1}} = 0) $ and $ p_{t_{n-1}}^{(2)} = P(X_{t_{n-1}} = 1) $, the corresponding vector at time $t_{n}$ is given by 
\begin{eqnarray}
p_{t_{n}} & = & P^{\top}p_{t_{n-1}}.
\label{eq: HMMHiddenRecursion}
\end{eqnarray}
In our example, the emission probabilities of $Y$ are given by a binary symmetric channel (random bit flip) with error probability $0<\delta<1$
\begin{eqnarray}
p(Y_{t_n}=1|X_{t_{n}}=0)=\delta, \quad p(Y_{t_n}=0|X_{t_{n}}=1)=\delta.
\end{eqnarray}
The structure of this model is depicted in Figure~\ref{fig:HMM}a.

For filtering, we can directly apply Eq.~\eqref{eq:VfS - filtering Bayes}, and since the state space is discrete, the integral reduces to a sum over the possible states $ 0 $ and $ 1 $.
Thus, Eq.~\eqref{eq:VfS - filtering Bayes} may be expressed as
\begin{eqnarray}
p(X_{t_n}|\Y_{t_0:t_n})  & =: & \hat{p}_{t_n} = \frac{\text{diag}(e(Y_{t_n})) P^{\top}\hat p_{t_{n-1}}}{e(Y_{t_n})^{\top}P^{\top}\hat p_{t_{n-1}}},
\label{eq: HMMFilterRecursion}
\end{eqnarray}
where $\text{diag}(v)$ is a diagonal matrix with the vector $v$ along the diagonal and $e$ is a vector encoding the emission likelihood:
\begin{eqnarray}
e(Y_{t_n}) &=&
\begin{pmatrix}
p(Y_{t_n}|X_{t_n}=0)\\
p(Y_{t_n}|X_{t_n}=1)
\end{pmatrix}.
\end{eqnarray}

Figure \ref{fig:HMM}b shows a sample trajectory of the hidden state $X_{t_n}$, the corresponding observations $Y_{t_n}$ as well as the filtered probabilities $p_{t_n}$ and the estimated state $\hat{X}_{t_n}$.
Even though what is presented here is a very simple setting (discrete time finite number of states), it illustrates nicely that the filter takes into account both the dynamics of the hidden states as well as the reliability of the observations.

\subsection{Continuous (state) space}
\label{sec:VfS - Continuous state space}

Remarkably, in the previous example the filtering problem could be solved in closed form because it was formulated in discrete time for a discrete state space.
We will now continue our journey towards more complex filtering problems involving a continuous state space.
For this, we have to go back to Eq.~\eqref{eq:VfS - filtering Bayes}, which is actually the filtering recursion for any state space.
While being straightforward to write down - is it possible to solve it in closed form?
Depending on the specific transition and emission densities, the integrals in Eq.~\eqref{eq:VfS - filtering Bayes} might not admit a closed-form solution.
In fact, this is almost always the case!
Except...


\subsubsection{The Kalman filter}

... if the transition and emission probabilities are Gaussians and linear, i.e.
\begin{eqnarray}
p(X_{t_n} | X_{t_{n-1}} ) & = & \mathcal{N} \left( X_{t_n}; A X_{t_{n-1}} , \Sigma_x \right) \label{eq:VfS - Kalman xt},
\\
p(Y_{t_n} | X_{t_{n}} ) & = & \mathcal{N} \left( X_{t_n}; B X_{t_{n}} , \Sigma_y \right) \label{eq:VfS - Kalman yt},
\end{eqnarray}
where we consider $ X_{t_n} \in \R^k $ and $ Y_{t_n} \in \R^l $ to be vector-valued random processes.
Further, $ A \in \R^{k \times k} $ and $ B \in \R^{l \times k} $ are the transition and emission matrices, respectively, and $ \Sigma_x \in \R^{n \times n} $  and $ \Sigma_y \in \R^{l \times l} $ are state and observation noise covariances, respectively.

Let us assume that at time $ t_{n-1} $ the posterior is given by a Gaussian
\begin{eqnarray}
p(X_\tnm|\Y_{0:{t_{n-1}}}) & = & \mathcal{N} (X_\tnm ; \mu_\tnm, \Sigma_\tnm).
\end{eqnarray}
We can immediately plug Eq.~\eqref{eq:VfS - Kalman xt} and \eqref{eq:VfS - Kalman yt} together with this assumption into Eq.~\eqref{eq:VfS - filtering Bayes}.
After a bit of tedious but straightforward algebra \citep[see][Section 13.3.1]{Bishop}, we find that the posterior is also a Gaussian $ \mathcal{N} ( X_t; \mu_{t_n} , \Sigma_{t_n} )$.
The famous Kalman filter equations give us update rules for its mean and variance:
\begin{eqnarray}
	\mu_{t_n} & = & A \mu_\tnm + K_t ( Y_{t_{n}} - B A \mu_\tnm ),
	\\
	\Sigma_{t_n} & = & (\mathbb{I} - K_t B) \tilde{\Sigma}_\tnm,
\end{eqnarray}
where $ \tilde{\Sigma}_\tnm = A \Sigma_\tnm A^\top + \Sigma_x $ is the variance of $ p(X_t|\Y_{0:{t_{n-1}}}) $, obtained after performing the marginalization over the state transition.
The so-called Kalman gain $ K_t $ is given by
\begin{eqnarray}
	K_t & = & \tilde{\Sigma}_\tnm B^\top ( B \tilde{\Sigma}_\tnm B^\top + \Sigma_y )^{-1}.
\end{eqnarray}
The immediate implication of this result is that for this particular model, given that the initial distribution is a Gaussian, the posterior stays Gaussian at all times.

\subsubsection{Particle filtering in discrete time}
\label{sec:Intro Filt - Particle filtering in discrete time}

In those cases where transition and emission probabilities are not Gaussian, we cannot expect Eq.~\eqref{eq:VfS - filtering Bayes} to take an analytically accessible form.
In other words:
as time goes by (in terms of time steps $ n $), we will have to keep track of an ever-growing amount of integrals, which is clearly not desirable.
Alternatively, we can try to approach this task numerically, by considering empirical samples and propagating these samples through time to keep track of this posterior.
This idea is the very basis of particle filters (PF).

The only remaining problem is that direct sampling from the true posterior is usually not possible.
In Section \ref{sec:Importance sampling} we have motivated importance sampling for a static setting from a change of measure perspective, and we will now use the same reasoning to motivate \emph{sequential} importance sampling.
In other words: we will replace samples from the true posterior (under $ \P $) by weighted samples from a proposal density under $ \Q $.
Importantly, a `sample' $ i $ here refers a single realization of the whole path $ \X_{0:t_n} = \{ X_0 , \dots , X_{t_n} \}$, and the measure change needs to be done with respect to the whole sequence of state and observations.

Let us first note that the posterior expectation can be understood as an expectation with respect to the whole sequence
\begin{eqnarray}
	\E_\P \left[ \phi(X_{t_n}) | \Y_{0 : t_n} \right] & = & 
	\int_S dx_{t_n} \, \phi(x_{t_n}) p(x_{t_n} | \Y_{0:t_n}) \nonumber
	\\
	& = &
	\int_S dx_{0:t_n} \, \phi(x_{t_n}) p( x_{0:t_n}  | \Y_{0:t_n}),
\end{eqnarray}
where in the last step we simply used that $ \int_S dx_{0:t_{n-1}} \, p( x_{0:t_{n-1}}  | \Y_{0:t_n}) =1$.
Now, we perform the measure change according to Eq.~\eqref{eq:Intro Filt - Bayes' rule exp change of measure} :
\begin{eqnarray}
	\E_\P \left[ \phi(X_{t_n}) | \Y_{0 : t_n} \right] & = & \frac{ \E_\Q [ L( \X_{0:t_n} ,\Y_{0:t_n} ) \phi(X_{t_n}) | \Y_{0 : t_n}  ] }{\E_\Q [ L( \X_{0:t_n} ,\Y_{0:t_n} ) | \Y_{0 : t_n}   ]},
\end{eqnarray}
with 
\begin{eqnarray}
	L( x_{0:t_n} ,y_{0:t_n} ) & = & \frac{ p(x_{0:t_n}, y_{0:t_n})  }{ q (x_{0:t_n}, y_{0:t_n}) } = \frac{ p(x_{0:t_n}| y_{0:t_n}) p ( y_{0:t_n}) }{ q (x_{0:t_n}| y_{0:t_n}) q ( y_{0:t_n})  }  , \label{eq:VfS - Radon Nikodym PF}
\end{eqnarray}
where $ p $ and $ q $ denote densities of $ \P $ and $ \Q $, respectively. 
Let us now choose the measure $ \Q $ such that the conditional density $ q ( x_{0:t_n} | y_{0:t_n} ) $ factorizes, i.e.
\begin{eqnarray}
	q ( x_{0:t_n} | y_{0:t_n} ) & = & \prod_{j=0}^n \pi ( x_{t_j} | x_{0:t_{j-1}} , y_{0:t_{j}}  ) \nonumber
	\\
	&=& \pi ( x_{t_n} | x_{0:t_{n-1}} , y_{0:t_{n}}  ) q ( x_{0:t_{n-1|}} | y_{0:t_{n-1}} ) . \label{eq:VfS - proposal particles}
\end{eqnarray}
Further, we can rewrite the conditional density $ p ( x_{0:t_n} | y_{0:t_n} ) $ using the structure of the SSM
\begin{eqnarray}
	p ( x_{0:t_n} | y_{0:t_n} ) & = & \frac{p(y_{t_n} | x_{0:t_n} , y_{0:t_{n-1}}) p(x_{0:t_n} | y_{0:t_{n-1}}) }{p(y_{t_n}|y_{0:t_{n-1}})} \nonumber
	\\
	& = & \frac{p(y_{t_n} | x_{0:t_n} , y_{0:t_{n-1}}) p(x_{t_n} | x_{0:t_{n-1}}, y_{0:t_{n-1}}) }{p(y_{t_n}|y_{0:t_{n-1}})} p ( x_{0:t_{n-1}} | y_{0:t_{n-1}} )  \nonumber
	\\
	& = & \frac{p(y_{t_n} | x_{t_n} ) p(x_{t_n} | x_{t_{n-1}} ) }{p(y_{t_n}|y_{0:t_{n-1}})} p ( x_{0:t_{n-1}} | y_{0:t_{n-1}} ) .
\end{eqnarray}
Thus, using  that all factors independent of the state variable can be taken out of the expectations in Eq.~\eqref{eq:VfS - Radon Nikodym PF} and cancel subsequently, we find
\begin{eqnarray}
L( x_{0:t_n} ,y_{0:t_n} ) & \propto & \frac{p(y_{t_n} | x_{t_n} ) p(x_{t_n} | x_{t_{n-1}} ) }{ \pi ( x_{t_n} | x_{0:t_{n-1}} , y_{0:t_{n}}  )  } \frac{p ( x_{0:t_{n-1}} | y_{0:t_{n-1}} )}{q ( x_{0:t_{n-1|}} | y_{0:t_{n-1}} )}  \nonumber
\\
&  \propto &\frac{p(y_{t_n} | x_{t_n} ) p(x_{t_n} | x_{t_{n-1}} ) }{ \pi ( x_{t_n} | x_{0:t_{n-1}} , y_{0:t_{n}}  )  } L( x_{0:t_{n-1}} ,y_{0:t_{n-1}} ). \label{eq:VfS - Radon Nikodym PF recursive}
\end{eqnarray}

In analogy to Section \ref{sec:Importance sampling}, we now take $ M $ i.i.d.~samples from the proposal density, i.e. we sample $ X^{i}_{0:t_n} \sim  q ( X_{0:t_n} | Y_{0:t_n} )  $, and weigh them according to the value of the likelihood ratio evaluated at the particle positions (cf.~Eq.~\ref{eq:VfS - PF static}).
Since the proposal in Eq.~\eqref{eq:VfS - proposal particles} was chosen to factorize, both the sampling process as well as the evaluation of the unnormalized importance weights  $ w_{t_n}^{(i)} $ (according to Eq.~\ref{eq:VfS - Radon Nikodym PF recursive}) can be done recursively.
More specifically, the problem of sampling (and weighing) the whole sequences $ \X^{(i)}_{0:t_n} $ is replaced by sampling just a single transition $ X_{t_n}^{(i)} $ for each of the $ M $ particles at each time step $ n $ and updating the associated particle weights.
\begin{eqnarray}
X_{t_n}^{(i)} & \sim & \pi (X_{t_n} | \X_{0:t_{n-1}}^{(i)}, \Y_{0:t_{n}}), \label{eq:Intro Filt - PF proposal} \\
w_{t_n}^{(i)} & = & L( \X_{0:t_n}^{(i)}, \Y_{0:t_{n}} ) = w_{t_{n-1}}^{(i)} \frac{p(Y_{t_n}|X_{t_n}^{(i)})\, p(X_{t_n}^{(i)}|X_{t_{n-1}}^{(i)} ) }{\pi (X_{t_n}^{(i)} | \X_{0:t_{n-1}}^{(i)}, \Y_{0:t_n})}, \label{eq:Intro Filt - PF weights unnormalized}
\end{eqnarray}
such that the posterior expectation is approximated by
\begin{eqnarray}
\E_\P \left[ \phi(X_{t_n}) | \Y_{0 : t_n} \right] & = & \frac{1}{Z_{t_n}} \sum_{i=1}^M w_{t_n}^{(i)} \phi(X_{t_n}^{(i)}),
\end{eqnarray}
with $Z_{t_n} = \sum_{i=1}^P w_{t_n}^{(i)}$.

A simple (but not necessarily efficient) choice is to use the transition probability $ p(X_{t_n} | X_{t_{n-1}} ) $ as the proposal function in Eq.~\eqref{eq:Intro Filt - PF proposal}.
Then, computation of the unnormalized weights simplifies to
\begin{eqnarray}
\tilde{w}_{t_n}^{(i)} & = & w_{t_{n-1}}^{(i)} p(Y_{t_n}|X_{t_n}^{(i)}). \label{eq:Intro Filt - bootstrap weights unnormalized}
\end{eqnarray}
This scheme is the basis of the famous \textbf{Bootstrap PF} (BPF, \citealp{Gordon1993})\footnote{Although technically, the BPF requires a resampling step at every iteration step.}.
\citet{Doucet2000} state that the BPF is ``inefficient in simulations as the state space is explored without any knowledge of the observations''.
To account for this, alternative proposal densities can be crafted in discrete time, which may take into account the observations in the particle transitions (e.g.~the `optimal proposal' in \citealp{Doucet2000}).

\section{Knowing where your towel is: setting the stage for continuous-time models}
\label{Sect2}

\begin{flushright}
\textit{``A towel is about the most massively useful thing an interstellar
hitchhiker can have.\\
Partly it has great practical value.
More importantly,\\a towel has immense psychological value.''}\\
--- Douglas Adams
\end{flushright}

So far, we have made our journey from Bayes' theorem to discrete-time filtering first for discrete state spaces and then made the transition towards continuous state space models.
The next logical step would be the transition to continuous time models.
In the following three sections, we will see that the mindset is very similar to the approaches taken before, just in their respective continuous-time limit, i.e.~$ dt = t_n - t_{n-1} \to 0 $.
In particular, we will use the change of measure approach to derive the filtering equations, i.e.~dynamical equations for the posterior expectations $ \E[\phi(X_t) | \Y_{0:t}] $ or, equivalently, the posterior density $ p(X_t|\Y_{0:t}) $.

But let us take a step back here and first explain the model assumptions under which we will present continuous-time filtering theory.
For the purpose of this tutorial, we have seen that a generative model consists of two parts:
\begin{enumerate}
    \item 
    A \emph{signal} model or \emph{hidden process} model that describes the dynamics of some system whose states we want to estimate.
   In continuous-time, we will consider two very general classes of signal model, namely continuous-time Markov chains (countable or finite state space) and jump-diffusion processes (continuous state space).
    \item 
    An \emph{observation model} that describes how the system generates the information that we can observe and utilize in order to estimate the state.
    We will elaborate the filtering theory for two types of observation noise, namely continuous-time Gaussian noise and Poisson noise.
\end{enumerate}

\subsection{Signal models}
As in Section \ref{sec:VfS}, we will restrict ourself to the treatment of Markovian processes for the signal, i.e.~$ p(X_t|X_{0:t-dt}) =  p(X_t|X_{t-dt}) $.
Our goal in this subsection will be to obtain dynamical equations that fully describe the signal process.

\subsubsection{Markov chain}
An important example is when $X_t$ is a continuous-time time-homogeneous Markov chain with a finite number of states, i.e.~$S=\{1,...,m\}$. 
In this case we may represent the function $\phi:\{1,...,m\} \to \mathbb{R}$ as a vector $\phi=(\phi(1),...,\phi(m))^{\top}$ and we have a transition probability matrix $P(t)^{\top}$. 
The entry $P_{ji}(t)$ gives the probability of going from state $j$ to state $i$ within a time interval of length $t$, so it is a time-dependent generalization of Eq.~\eqref{eq:HMM transition}.
This allows us to compute the distribution at time $t$, $p(t)$ from the initial distribution $p(0)$ as $p(t)=P(t)^{\top}p(0)$.
We therefore have two equivalent ways of computing the expectation
of $\phi$:
\begin{eqnarray}
    \mathbb{E}[\phi(X_t)] & = &  p(t)^{\top}\phi \nonumber
    \\
    &= & p(0)^{\top} P(t) \phi = p(0)^T \phi(t). \label{eq:towel - Markov chain expectation}
\end{eqnarray}
In the first one, the observable is fixed while the distribution changes as a function of time, while in the second, the distribution is fixed to the initial distribution, and the observable evolves in time, i.e. $\phi(t)=P(t)\phi$.

By differentiating with respect to time, we obtain differential equations for the distribution $p(t)$ and the observable $\phi(t)$,
\begin{eqnarray}
\dot{\phi}(t)=\dot{P}(t)\phi, \label{eq:MC - time evolution observable}\\
\dot{p}(t)=\dot{P}(t)^{\top}p(0).
\end{eqnarray}
The Markov property ensures that $P(t+s)=P(t)P(s) = P(s)P(t)$.
Further, since $ P(0) = \mathbb{I} $ is the unit matrix\footnote{Because $ p(t=0) = P(0)^\top p(0) $ is only fullfilled if $P(0) = \mathbb{I} $.}, the time derivative of the matrix $P(t)$ can be simplified to
\begin{eqnarray}
\dot{P}(t) & = & \lim_{s\to 0} \frac{P(t+s) - P(t)}{s} = P(t) \lim_{s\to 0} \frac{P(s) - \mathbb{I}}{s} \nonumber
\\
& = & P(t) \dot{P} (0).
\end{eqnarray}
We denote $A=\dot P(0)$ and then get
\begin{eqnarray}
\dot{\phi}(t)=A\phi(t),\\
\dot{p}(t)=A^{\top}p(t). \label{eq:Intro Filt - AdjointEqMC}
\end{eqnarray}
Equivalently, we find for the time derivative of the expectation
\begin{eqnarray}
	\frac{d}{dt}\E [ \phi(X_t) ] & = & p(0)^\top \dot{P}(t) \phi  = p(t)^\top A \phi = \E [A \phi ]. \label{eq:Intro Filt - PostEqMC}
\end{eqnarray}
So conceptually, the whole temporal evolution of the stochastic process $ X_t $ is encapsulated in the matrix $ A $, the so-called \emph{generator} matrix.
In other words, the generator matrix, together with the initial distribution, is all we need to completely characterize the Markov chain process.

\subsubsection{Jump-diffusion process}

Intuitively, in order to make the transition to a continuous state space, we have to exchange ``sums by integrals and matrices by linear operators''.
We will now see that this holds for the hidden state dynamics by characterizing a continuous-time stochastic process with continuous state space $ S $ similarly to the equations \eqref{eq:Intro Filt - AdjointEqMC} and \eqref{eq:Intro Filt - PostEqMC} above.

\begin{figure}
  \centering
      \includegraphics[width=\textwidth]{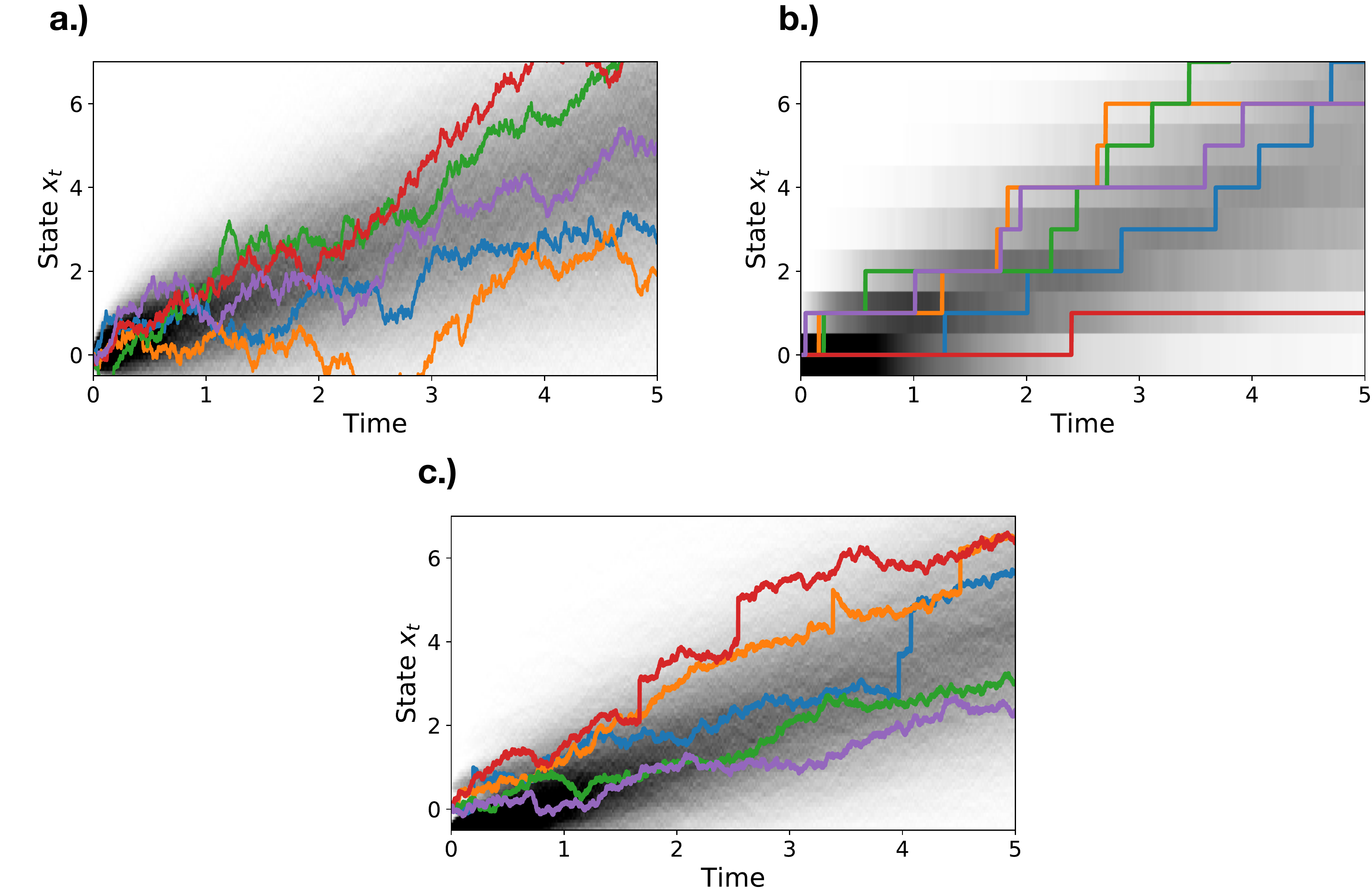}
  \caption{ Example trajectories from Eq.~\eqref{eq:Intro Filt - JumpDiffSignal}.
  Shading denotes density of 10'000 simulated trajectories.
  a.) Drift diffusion process ($f(x)=1, G(x)=1,J(x)=0$). b.) Jump process ($f(x)=0, G(x)=0,J(x)=1$) with rate $\lambda(x) = 1$. c.) Jump diffusion process ($f(x)=1/2, G(x)=1/2,J(x)=1$) with rate $\lambda(x) = 1/2$.
   \label{fig:JumpDiff}}
\end{figure}

An important signal model, which is a generalization of the classical diffusion model in continuous time, is a hidden state $X_t$ that is a jump-diffusion process, i.e. it evolves according to a stochastic differential equation (SDE) in $S=\mathbb{R}^n$,
\begin{eqnarray}
dX_t &=& f(X_t,t)\,dt+G(X_t,t)\, dW_t+ J(X_t,t)dN_t. \label{eq:Intro Filt - JumpDiffSignal}
\end{eqnarray}
Here, $f:\mathbb{R}^n \times \mathbb{R} \to \mathbb{R}^n$, $G :  \mathbb{R}^n \times \mathbb{R} \to \mathbb{R}^{n\times n}$, and $J: \mathbb{R}^n \times \mathbb{R} \to \mathbb{R}^{n\times k}$ are called the drift, diffusion, and jump coefficients of $X_t$, respectively.
The process noise is modelled by two types of noise sources: $ W_t \in \R^n $ is a vector Brownian motion that models white Gaussian noise in continuous time, and we may consider $ d W_t \sim  \mathcal{N}(0,dt) $.
$N_t$ is a $k$-dimensional point process with $k$-dimensional rate (or intensity) vector $\lambda(X_t)$, i.e.~$  	dN_t  \sim  \text{Poisson} ( \lambda (X_t) dt ) $.
Note that $ dN_t $ takes only values 0 or 1, because in the limit $ dt \to 0 $, the Poisson distribution becomes a Bernoulli distribution.
In Figure \ref{fig:JumpDiff}, we show example trajectories from Eq.~\eqref{eq:Intro Filt - JumpDiffSignal}, one being a drift-diffusion (where the jump term vanishes), one being a pure jump process, and the last one being a jump-diffusion process.


Dealing with this type of SDE model is considerably more technical than the Markov chains above.
Therefore, we will outline the theory of diffusion processes for readers who are new to them.
Unless stated otherwise, derivations presented here roughly follow \citet{Gardiner} and \citet{bremaud1981point}.

We can choose to describe the process in terms of transition probability densities $p(x,t|x',s)$, which give the probability density at a point $x\in\mathbb{R}^n$ at time $t$ conditioned on starting at a point $x'\in\mathbb{R}^n$ at time $s<t$.
This transition density can be combined with the initial density $p_0(x')$ by integrating in order to compute an expectation:
\begin{eqnarray}
\mathbb{E}[\phi(X_t)]&=&\iint  \phi(x) p(x,t|x',0)p_0(y) \,dx \, dx' \nonumber \\
&=&\int \phi(x) p(x,t) \, dx,
\end{eqnarray}
in complete analogy with the vector-matrix-vector product for the Markov chains in Eq.~\eqref{eq:towel - Markov chain expectation}.
Taking this analogy further, differentiating with respect to time gives rise to two different (equivalent) ways of writing the time evolution of the expected value:
\begin{eqnarray}
\frac{d}{dt}\mathbb{E}[\phi(X_t)] &=&\int\phi(x)\partial_tp(x,t) \, dx\nonumber \\
&=&\int\phi(x) \mathcal{A}^{\dag}p(x,t) \, dx \nonumber \\
&=&\int \mathcal{A}\phi(x) p(x,t) \, dx,
\label{eq:Intro Filt - IntegrationByParts}
\end{eqnarray}
where in analogy to Eq.~\eqref{eq:Intro Filt - PostEqMC} we have introduced the adjoint operator $\mathcal{A}^{\dag}$ that describes the time evolution of the probability density.
Thus, in analogy to Eq.~\eqref{eq:MC - time evolution observable} we can set out to find the appropriate form of the generator $\mathcal{A}$, which generalizes the generator matrix $ A $ of the Markov chain, and then, by integration by parts, we may derive the corresponding adjoint operator $\mathcal{A}^{\dag}$.

%

\paragraph{It\^{o} lemma for jump diffusions}

The form of the generator $ \mathcal{A} $ can be obtained by changing the variables in Eq.~\eqref{eq:Intro Filt - JumpDiffSignal} from the random variable $ X_t $ to the random variable $ \phi_t := \phi(X_t) $. 
The following calculation will be performed for a scalar process $ X_t $.\footnote{Generalization to a multivariate state process $ X_t $ is straightforward.}
Consider an infinite Taylor expansion of its increment $ d\phi_t $ around $ dX_t = 0 $ up to $ \mathcal{O}(dt) $:
\begin{eqnarray}
	d \phi_t & = & \phi(X_t + dX_t) - \phi(X_t) \nonumber 
	\\
	& = & \sum_{n=1}^\infty \frac{1}{n!}  \phi_t^{(n)}  \, (dX_t)^n \label{eq: Appendix - Ito expansion},
\end{eqnarray}
with $ \phi_t^{(n)}:= \left( \partial_x^n \phi(x) \right)|_{x=X_t} $.

In a deterministic differential, Taylor-expanding up to first order would suffice since $ dt^n=0 $ $ \forall n>1 $.
In Eq.~\eqref{eq:Intro Filt - JumpDiffSignal}, the additional stochastic terms add additional orders of $dt$.
Particularly, since the variance of the Brownian motion process grows linearly in time, we have $dW_t^2 = dt$, and thus the diffusion term has to be expanded up to second order.
For the jump term, all order up to infinity have to be considered:  $ N_{t} $ is not a continuous process, with jumps of always size 1 irrespectively of the infinitesimally small time interval $ dt $.
Therefore, any power of this jump will  have the same magnitude, i.e.~$  dN_t^n = d N_{t} $, $ \forall n $.
Thus, we find for a scalar process
\begin{equation}
\begin{split}
d \phi_t & =  \left[ f(X_t,t)\phi_t'+ \frac{1}{2} G^2(X_t,t)\phi_t''  \right] dt + G(X_t,t)\phi_t'\, dW_t  \\
&\quad + \sum_{n=1}^\infty \frac{1}{n!} J^n(X_t,t) \phi_t^{(n)} \, dN_t
\\
& =  \left[ f(X_t,t)\phi_t'+ \frac{1}{2} G^2(X_t,t)\phi_t''  \right] dt + \left[ \phi \left( X_t + J(X_t) \right) - \phi(X_t) \right] \, \lambda(X_t)dt \\
& \quad + G(X_t,t)\phi_t'\, dW_t + \left[ \phi \left( X_t + J(X_t) \right) - \phi(X_t) \right] \, \big(dN_t-\lambda(X_t)dt\big) 
\\
& =:  \mathcal{A}\phi_t \, dt + dM_t^\phi. 
\end{split} \label{eq:Appendix - Ito lemma}
\end{equation}
This formula is called \textbf{It\^{o} lemma}.
In the last step, we have defined
\begin{eqnarray}
	\mathcal{A} \phi_t & = & f(X_t,t)\phi_t'+ \frac{1}{2} G^2(X_t,t)\phi_t'' +  \lambda(X_t) ( \phi \left( X_t + J(X_t) \right) - \phi(X_t) ), \label{eq:Intro Filt - InfinitesimalGenerator}
	\\
	 dM_t^\phi & = & G(X_t,t)\phi_t'\, dW_t + \left[ \phi \left( X_t + J(X_t) \right) - \phi(X_t) \right] \, \big(dN_t-\lambda(X_t)dt\big),
\end{eqnarray}
where $ \mathcal{A} $ is the infinitesimal \textbf{generator} of the stochastic process $ X_t $.
The stochastic process $ M_t^\phi $ is a so-called martingale\footnote{Loosely speaking, a martingale is a sequence of random variables, whose conditional expectation in the next time step is equal to the value of the random variable at the current time step.} and the contribution from its increment vanishes upon taking expectations, i.e.~$ \E[ M_t^\phi ] = 0 $.
Thus, taking expectations on both sides of Eq.~\eqref{eq:Appendix - Ito lemma} we find indeed
\begin{eqnarray}
\frac{d}{dt}\mathbb{E}[\phi(X_t)]=\mathbb{E}[\mathcal{A}\phi(X_t)],
\label{eq:Intro Filt - InfinitesimalGenerator expectation}
\end{eqnarray}
which is the continuous state space analogue to Eq.~\eqref{eq:Intro Filt - PostEqMC}.

The multivariate version is completely analogous:
\begin{equation}
d\phi(X_t)=\mathcal{A}\phi(X_t)dt+dM_t^{\phi}, \nonumber
\end{equation}
where now the infinitesimal generator of the stochastic process is given by
\begin{equation}
\begin{split}
\mathcal{A}\phi(x) &= \sum_{i=1}^n f_i(x,t)\partial_{x_i} \phi(x)+\frac{1}{2}\sum_{i,j=1}^n \big(G G^{\top}(x,t)\big)_{ij}\partial_{x_i}\partial_{x_j}\phi(x)\\
&\quad +\sum_{i=1}^k\lambda_i(x)\Big[\phi\big(x+J_i(x,t)\big)-\phi(x)\Big ],
\end{split}
\label{eq:Intro Filt - InfinitesimalGeneratorDiffusion}
\end{equation}
and the martingale part reads
\begin{equation}
\begin{split}
dM^{\phi}_t&=\sum_{i,j=1}^n  G_{ij}(X_t,t) (\partial_{x_i}\phi(x) |_{x=X_t}) dW^{j}_s \\
&\quad +\sum_{i=1}^k   \left[ \phi \left( X_t + J_i(X_t) \right) - \phi(X_t) \right] \, \big(dN^{i}_t-\lambda_i(X_t)ds\big).
\end{split}
\label{eq: Appendix - MartingaleMultivariate}
\end{equation}
The operator $\mathcal{A}$, just like the generator matrix $ A $ of the Markov chain, together with the initial distribution, completely characterizes the Markov process and allows us to describe its time evolution on an abstract level.
Or in other words: even though the particular form of $\mathcal{A}$ might be different for each of these models presented here, the structure of the mathematics remains the same, and can therefore be generalized to arbitrary $\mathcal{A}$ when the need arises.




\paragraph{The evolution of the probability density}

\label{sec:Appendix - The evolution of the probability density}
With the explicit form in Eq.~\eqref{eq:Intro Filt - InfinitesimalGeneratorDiffusion} of the generator $\mathcal{A}$, we can go back to Eq.~\eqref{eq:Intro Filt - IntegrationByParts} and perform the integration by parts to find the adjoint operator $\mathcal{A}^{\dag}$, which will take the role of $A^{\top}$ in the Markov chain case.

Plugging Eq.~\eqref{eq:Intro Filt - InfinitesimalGeneratorDiffusion} into Eq.~\eqref{eq:Intro Filt - IntegrationByParts}, we obtain
\begin{eqnarray}
\int_{\mathbb{R}^n} \mathcal{A}\phi(x) p(x,t)dx & = & \sum_{i=1}^n\int_{\mathbb{R}^n} f(x,t)_i\partial_{x_i}\phi(x)p(x,t) dx \nonumber\\
&&+\frac{1}{2}\sum_{i=1}^n\sum_{j=1}^m\int_{\mathbb{R}^n} \big(G G^{\top}(x,t)\big)_{ij}\partial_{x_i}\partial_{x_j}\phi(x)p(x,t)dx \nonumber \\
&&+\sum_{i=1}^k\int_{\mathbb{R}^n}\lambda_i(x)\Big[\phi\big(x+J_i(x,t)\big)-\phi(x)\Big ]p(x,t)dx. \label{eq:FPE - integrals}
\end{eqnarray}
The first two integrals can be dealt with by ordinary integration by parts\footnote{Here, we make the assumption that the density $p(x,t)$ and all its derivatives vanish at infinity.}, i.e.
\begin{eqnarray}
\int_{\mathbb{R}^n} f(x,t)_i\partial_{x_i}\phi(x)p(x,t) dx = -\int_{\mathbb{R}^n} \phi(x)\big[\partial_{x_i}f(x,t)_ip(x,t)\big] dx,
\end{eqnarray}
and
\begin{eqnarray}
\int_{\mathbb{R}^n} \big(G G^{\top}(x,t)\big)_{ij}\partial_{x_i}\partial_{x_j}\phi(x)p(x,t)dx=\int_{\mathbb{R}^n}\phi(x)\partial_{x_i}\partial_{x_j}\Big[\big( G G^{\top}(x,t)\big)_{ij}p(x,t)\Big]dx.
\end{eqnarray}
For the third integral in Eq.~\eqref{eq:FPE - integrals}, we perform a change of variables $x\mapsto x+J_i(x,t)$ (where we assume the integral boundaries to not be affected by the substitution), thereby obtaining 
\begin{eqnarray}
&\int_{\mathbb{R}^n}\lambda_i(x)\Big[\phi\big(x+J_i(x,t)\big)-\phi(x)\Big ]p(x,t)dx \nonumber \\ &=\int_{\mathbb{R}^n}\phi(x)\Big[\lambda_i(x-J_i(x,t))p(x-J_i(x,t),t) \det{\frac{\partial( J_{1i}, \dots , J_{ji} ) }{\partial (x_1, \dots , x_j)}}  -\lambda_i(x)p(x,t)\Big ]dx,
\end{eqnarray}
where $ \det{\frac{\partial( J_{1i}, \dots , J_{ji} ) }{\partial (x_1, \dots , x_j)}} $ denotes the Jacobi determinant of the entries of the column vector $ J_i $.
Combining all of these integrals (including the sums) and comparing with Eq.~\eqref{eq:Intro Filt - IntegrationByParts}, we can therefore read off the form of the adjoint operator:
\begin{equation}
\begin{split}
\mathcal{A}^{\dag}p(x,t) &= \sum_{i=1}^n \partial_{x_i}\Big[f(x,t)_ip(x,t)\Big]+\frac{1}{2}\sum_{i,j=1}^n\partial_{x_i}\partial_{x_j}\Big[\big(G G^{\top}(x,t)\big)_{ij}p(x,t)\Big]\\
&\quad +\sum_{i=1}^k\Big[\lambda_i(x-J_i(x,t))p\big(x-J_i(x,t),t\big)  \det{\frac{\partial( J_{1i}, \dots , J_{ji} ) }{\partial (x_1, \dots , x_j)}} -\lambda_i(x)p(x,t)\Big ].
\end{split}
\end{equation}
Using Eq.~\eqref{eq:Intro Filt - IntegrationByParts} once more, we find the evolution equation for the density $p(x,t)$:
\begin{eqnarray}
	\partial_tp(x,t) & = & \mathcal{A}^{\dag}p(x,t). \label{eq:Appendix - Fokker Planck equation}
\end{eqnarray}
If we leave out the jump terms, this is called the \textbf{Fokker-Planck equation} or Kolmogorov forward equation.
With the jump terms, it is often referred to as the \textbf{Master equation}.

\subsection{Observation model}

In the previous section, we have encountered various signal models, which are the processes we want to infer.
The knowledge about how these processes evolve in time, formally given by the generator $ \mathcal{A} $, serves as prior knowledge to the inference task.
Equally important, we need measurements, or observations, to update this prior knowledge.
In particular, an observation model describes how the signal gets corrupted during measurement.
This may comprise both a lossy transformation (e.g. only certain components of a vector-valued process are observed), and some stochastic additive noise that randomly corrupts the measurements.
Roughly speaking, the measurements $Y_t$ are given by 
$$Y_t=h(X_t)+\text{noise},$$
but we will need to be careful about the precise way in which the noise is added in order to make sense in continuous time.

In the following, we will consider two types of noise: Gaussian and Poisson.
The simplicity of noise of these two noise models greatly simplifies the formal treatment of the filtering problem, and while the two types of noise seem very different, there is a common structure that will emerge.

When considering more general noise models than the ones below, the technique of Section~\ref{Sect3} (change of measure) can be applied whenever the observation noise (whatever is added to the deterministic transformation) is additive and independent of the hidden state.

\subsubsection{Continuous-time Gaussian noise}
\label{sec:GM - Obs - Cont Gaussian}
The simplest noise model is often white Gaussian noise.
For continuous-time observations, however, one cannot simply take an observation model $Y_t=h(X_t)+\eta_t$ with independent Gaussian $\eta_t$ because for a reasonably well-behaved process $X_t$, an integration of $Y_t$ over a finite time interval would completely average out the noise and therefore allow one to perfectly recover the transformed signal $h(X_t)$.\footnote{If the observations are made at discrete times $t_1,t_2,...$, this is not problematic. 
Filtering of a continuous-time hidden process with discrete-time observations is reviewed in \citet[p.~163ff]{Jazwinski1970}. 
If the observation model in Eq.~\eqref{eq:Intro Filt - GenerativeModel Observations} is discretized, one gets back to a discrete-time observation model with Gaussian noise.}
The filtering problem would therefore be reduced to simply inverting $h$.

One way of resolving the problem of finding a (nontrivial) model of white Gaussian noise is to switch to a differential form and use increments of the Wiener process as a noise term.
One therefore obtains an SDE for the \textbf{observation process} $Y_t$:
\begin{eqnarray}
dY_t &=& h(X_t,t)\,dt+\Sigma_y(t)^{1/2} \,dV_t. \label{eq:Intro Filt - GenerativeModel Observations}
\end{eqnarray}
Here, $  h: \mathbb{R}^n \times \mathbb{R} \to \mathbb{R}^l $ is a vector-valued function that links the hidden state (and time, if time-dependence is explicit) with the deterministic drift of the observations.
Further, $ V_t \in \R^l $ is a vector Brownian motion process and
$ \Sigma_y(t) : \mathbb{R} \to \mathbb{R}^{l\times l}$ is the time-dependent \emph{observation noise covariance}.

In the standard literature, one usually finds the special case $ \Sigma_y = \mathbb{I}^{m \times m} $, which is equivalent to Eq.~\eqref{eq:Intro Filt - GenerativeModel Observations} if the increment of the observation process $ Y_t $ is rescaled accordingly:
\begin{eqnarray}
	d\tilde{Y}_t &=& \Sigma_y(t)^{-1/2} d Y_t = \tilde{h}(X_t,t)\,dt+ dV_t, \label{eq:Intro Filt - GenerativeModel Observations rescaled}
\end{eqnarray}
where $ \tilde{h}(x,t) = \Sigma_y(t)^{-1/2} h(x,t) $ is the rescaled observation function.

\subsubsection{Poisson noise}
In many fields, observations come in the form of a series of events.
Examples include neuroscience (neural spike trains), geoscience (earthquakes, storms), financial transactions, etc.
This suggests a point process (whose output is a series of event times) or counting process (which counts the number of events) observation model.
A simple, but versatile, model for events is a Poisson process $ N_t $ with time-varying and possibly history-dependent intensity.
As an observation model, this doubly-stochastic Poisson process (also known as Cox process, \citealt{Cox1955}) has an intensity that depends on its own history as well as the hidden state.
We can describe this process by
\begin{eqnarray}
dN^{i}_t  &\sim&  \text{Poisson}\left( R^{i}_t\,dt\right), \quad i=1,..,l \label{eq:Intro Filt - Generative Model PP Observations}
\end{eqnarray}
where the intensity processes $R^{i}_t$ are nonnegative processes that can be computed from the current value of $X_t$ and the history of observations $\N_{0:s}$ for $s<t$.

To keep the notation simple, we will assume that the vector $R_t$ of intensities is given by a function of the hidden state,
\begin{eqnarray}
R_t=h(X_t),
\end{eqnarray}
but history-dependence in the form $R_t=h(X_t,\N_{0:t^-})$ does not significantly increase the difficulty of the filtering (given the observation, any history-dependence of the intensity is deterministic and can be factored out of the conditional expectation of the intensity).

%% file: 03_Filtering-formal.tex
\section{The answer to life, the universe and (not quite) everything: the filtering equations}
\label{Sect3}
\begin{flushright}
\textit{``I’ve just been created. I’m completely new to the
Universe in all respects.\\ Is there anything you can tell me?''}\\
--- Douglas Adams
\end{flushright}

The filtering problem is to compute the posterior (or conditional) density of the hidden state conditioned on the whole sequence of observations up to time $ t $, $ \Y_{0:t} $, or equivalently, to compute the posterior expectation (if it exists) of any real-valued measurable function $ \phi:\R^n\to\R $,
\begin{eqnarray}
\E_\Pmeasure \left[ \phi(X_t) | \Y_{0:t} \right] & = &\int_{-\infty}^\infty p (x|\Y_{0:t}) \phi(x) \, dx =: \Braket{\phi_t }_\Pmeasure, \label{eq:Intro Filt - posterior expectation definition}
\end{eqnarray}
where we use subscript $ \Pmeasure $ to indicate expectations with respect to the original probability measure $ \Pmeasure $.

That this is not an easy problem should be clear by now, because already the discrete-time filtering task (e.g.~in Eq.~\ref{eq:VfS - filtering Bayes}) involved a computation of as many integrals as there are time steps. 
In continuous time, this would amount to an infinite number of integrals.
This continuous-time problem has already been recognized and tackled by mathematicians in the 60s and 70s of the last century, providing formal solutions for the posterior density in terms of stochastic partial differential equations \citep{Kushner1964,Zakai1969}.
In the following, we will derive these equations, using what we have been using in the previous sections as our ultimate ``Point of View Gun for nonlinear filtering'':\footnote{The Point of View Gun is a weapon that causes its target to see things from the side of the shooter. Actually, it never appeared in any of Douglas Adams' novels, but it was featured in the 2005 movie.} the change of probability measure method.\footnote{There are other methods to arrive at the same equations, for instance the \emph{innovations approach} \citep[Chpt.~3.7]{Bain2009} or the more heuristic continuum limit approach originally taken by \citet{Kushner1964}.}

\subsection{Changes of probability measure - once again}

Let us once more revisit the change of probability measure in the context of filtering.
The goal is to pass from the original probability measure $\Pmeasure$ (under which the processes behave as our signal and observation model dictates), to an equivalent measure $\Qmeasure$, called \emph{reference measure}, under which the observation process becomes simpler and decouples from the signal process.
Here, we will finally generalize our introductory treatment from Section \ref{sec:Changes of measure} to stochastic processes.
Unsurprisingly, the calculations are quite similar.

If $\Pmeasure$ is a probability measure and we have a collection of processes ($X_t$ and $Y_t$), the measure $\Pmeasure_t$ is the restriction of $\Pmeasure$ to all events that can be described in terms of the behavior of $X_s$ and $Y_s$ for $0\leq s\leq t$.
If $ \P $ and $ \Q $ are equivalent, also their restrictions $ \P_t $ and $ \Q_t $ are equivalent.\footnote{
For stochastic processes, equivalence implies having the same noise covariance.}
The \textbf{Radon-Nikodym theorem} \citep[Theorem 10.6, p.~272ff]{Klebaner2005} then states that a random variable $ L_t $ exists, such that for all functions $\phi$
\begin{eqnarray}
	\E_\P \left[ \phi(X_t) \right] & = & \E_\Q \left[ L_t \cdot \phi(X_t) \right], \label{eq:Appendix - Radon Nikodym theorem}
\end{eqnarray}
where $ L_t = \frac{d \P_t}{d \Q_t} $ is called the Radon-Nikodym derivative or density of $\Pmeasure_t$ with respect to $\Qmeasure_t$. 
This is the generalization of Eq.~\eqref{HeuristicRN} in Section \ref{sec:Changes of measure}.

In analogy to Eq.~\eqref{eq:Appendix - Radon Nikodym theorem}, also the conditional expectations can then be rewritten in terms of a reference probability measure $ \Qmeasure $:
\begin{eqnarray}
\E_\Pmeasure \left[ \phi_t | \Y_{0:t} \right] & = & \frac{\E_\Qmeasure [\phi_t\, L_t | \Y_{0:t} ]}{\E_\Qmeasure [ L_t | \Y_{0:t} ]} = \frac{1}{Z_t} \Braket{\phi_t L_t }_\Qmeasure. \label{eq:Intro Filt - Kallianpur Striebel}
\end{eqnarray}
Equation \eqref{eq:Intro Filt - Kallianpur Striebel} is known as a Bayes' formula for stochastic processes (compare Eq.~\ref{eq:Intro Filt - Bayes' rule exp change of measure}) or \textbf{Kallianpur-Striebel formula}.
Here, we require a time-dependent normalization $ Z_t := \E_\Qmeasure [ L_t | \Y_{0:t} ]$, and $ \Braket{\phi_t L_t }_\Qmeasure := \E_\Qmeasure [\phi_t\, L_t | \Y_{0:t} ]  $ was introduced to keep the notation concise.
This generalizes Eq.~\eqref{eq:Intro Filt - Bayes' rule exp change of measure} above.

But wait: what exactly does the Radon Nikodym derivative $ L_t $ look like?
This really depends on the measure change we are about to perform, but it helps to recall that in a discrete-time (or actually already static) setting the equivalent of the Radon-Nikodym derivative is nothing else than the ratio between two probability distributions.
For the filtering problem below, we will choose a reference measure $ \Q $ such that the path of the observations $ \Y_{0 : t} $ (or equivalently the set of the increments $ dY_{0:t} $) becomes independent of the path of the state process $ \X_{0:t} $, i.e.~$ q(\X_{0:t},dY_{0:t}) = p(\X_{0:t}) q(dY_{0:t}) $.
This is very convenient, as this allows us to compute expectations with respect to the statistics of the state process (and we know how to do that).
Equation \eqref{eq:Intro Filt - Lratio} then directly tells us what the likelihood ratio has to look like for this measure change:
\begin{eqnarray}
	 L(x_{0:t},dy_{0:t}) &=& \frac{p(dy_{0:t}|x_{0:t})}{q(dy_{0:t})} = \frac{\prod_{s=0}^t p(dy_s|x_s)}{q(dy_{0:t})},
\end{eqnarray}
where now $ x_{0:t} $ and $ dy_{0:t} $ are variables reflecting the whole path of the random variable $ X_t $ and the set of infinitesimal increments $ dY_{0:t} $.
Importantly, this particular measure change is agnostic to how the hidden state variable $ X_t $ evolves in time, but just takes into account how the observations are generated via $ p(dy_t|x_t) $.

Let us first consider Gaussian observation noise, as encountered in Section \ref{sec:GM - Obs - Cont Gaussian}.
From Eq.~\eqref{eq:Intro Filt - GenerativeModel Observations}, we know that $ dY_t \sim \mathcal{N} ( dY_t ; h(X_t) dt, \Sigma_y dt) $.
Further, we choose $ q(dY_{t}) = \mathcal{N} (dY_t; 0,\Sigma_y dt) $ under the reference measure $ \Q $.
Thus, the Radon-Nikodym derivative $L_t = L(\X_{0:t},dY_{0 : t})$ can be written as
\begin{eqnarray}
	L_t & = &  \frac{\prod_{s=0}^t p(dY_s|X_s)}{ \prod_{s=0}^t q(dY_{s})} =  \prod_{s=0}^t \frac{\mathcal{N} ( dY_s; h(X_s) ds, \Sigma_y ds)}{\mathcal{N} (dY_s; 0,\Sigma_y ds)}   \nonumber
	\\
	& = & \prod_{s=0}^t \exp \left[  h(X_s)^\top \Sigma_y^{-1} dY_s - \frac{1}{2} h(X_s)^\top \Sigma_y^{-1} h(X_s) ds \right] \nonumber
	\\
	& \overset{\lim_{dt \to 0}}{=} & \exp \left[ \int_{0}^{t} \, h(X_s)^\top \Sigma_y^{-1} dY_s - \frac{1}{2} h(X_s)^\top \Sigma_y^{-1} h(X_s) ds \right], \label{eq:Intro Filt - RadonNikodym orig}
\end{eqnarray}
where in the last step we took the continuum limit $ \lim_{dt \to 0} $.
Consistently, we would have obtained this result if we had simply (and mindlessly) applied a theorem called Girsanov's theorem, choosing the reference measure $ \Q $ under which the rescaled observations process $ \tilde{Y}_t $ is a Brownian motion process (\citealp{Klebaner2005}, Chapter 10.3, on p.~274, in particular Remark 10.3).

Similarly, we can compute the Radon-Nikodym derivative for observations corrupted by Poisson noise as in Eq.~\eqref{eq:Intro Filt - Generative Model PP Observations}.
Here, we choose $ \Q $ such that $ q(dN_{0:t}) $ is Poisson with a constant reference rate $ \lambda_0 $.
The corresponding density reads
\begin{eqnarray}
	L_t & = & \prod_{s=0}^{t} \prod_{i=1}^{l} \frac{p(dN_s^i|X_{s})}{q(dN_s^i)} =  \prod_{s,i} \frac{\text{Poisson}(dN_s^i; h_i(X_s) ds)}{\text{Poisson}(dN_s^i; \lambda_0 ds)} \nonumber 
	\\
	& = & \prod_{s,i} \exp \left[   \lambda_0 ds -h_i(X_s) \, ds  + \log \frac{h_i(X_t)}{\lambda_0 }  dN_s^i \right] 
	\\
	& \overset{\lim_{dt \to 0}}{=} & \prod_{i=1}^l \exp \left[  \int_0^t (\lambda_0 -h_i(X_s)) \, ds + \log \frac{h_i(X_t)}{\lambda_0 }  dN_s^i \right].
\end{eqnarray}
Again, the same result could have been obtained with a Girsanov theorem \citep[see][Chapter VI, Theorems T2 and T3]{bremaud1981point}.

\subsection{Filtering equations for observations corrupted by Gaussian noise}

We are now equipped with the necessary tools to tackle the derivation of the filtering equations.
Here, the derivation will be briefly outlined \citep[for a more detailed and formal derivation, see][]{vanHandel2007,Bain2009}.

\label{sec:Intro Filt - Filtering problem continuous time}
As we stated in the beginning of this Section, we want to find a \emph{convenient} reference measure which decouples the signal and observations and at the same time turns the observations into something simple.
Recall that the Radon-Nikodym derivative (expressed for the rescaled observations process $ \tilde{Y}_t $) then takes the form
\begin{eqnarray}
L_t & = & \frac{d\P_t}{d\Q_t} = \exp \left[ \int_0^{t}  \tilde{h}(X_s)^{\top}   d \tilde{Y}_s - \frac{1}{2} \int_{0}^{t}  \norm{ \tilde{h}(X_s)}^2\, ds\right]. \label{eq:Intro Filt - RadonNikodym}
\end{eqnarray}
which evolves according to the following SDE (see \ref{sec:Appendix - RN evolution} for calculation steps):
\begin{eqnarray}
dL_t & = & L_t \tilde{h}(X_t)^{\top}  d \tilde{Y}_t. \label{eq:Intro Filt - RadonNikodym evolution}
\end{eqnarray}

Under $\Qmeasure$, the stochastic differential can be taken inside the expectation \citep[see][Chapter 7, Lemma 7.2.7]{vanHandel2007}, and we therefore obtain using It\^{o}'s lemma\footnote{Recall that this corresponds to a Taylor expansion up to second order (i.e.~first order in $dt$, since $\mathcal{O}(dW_t) = dt^{1/2}$) for diffusion processes, which is why we have to consider the product of differentials (product rule for stochastic differentials).}
\begin{eqnarray}
d\E_\P\left[ \phi_t | \Y_{0:t} \right] & = & d \left( \frac{1}{Z_t} \Braket{\phi_t L_t }_\Q \right) \nonumber 
\\
& = & \frac{1}{Z_t} \, d \Braket{\phi_t L_t }_\Q + \Braket{\phi_t L_t }_\Q \, d \left(\frac{1}{Z_t}\right) + d \Braket{\phi_t L_t }_\Q \, d \left( \frac{1}{Z_t}\right) \nonumber
\\
& = & \frac{1}{Z_t} \, \Braket{ d( \phi_t L_t ) }_\Q + \Braket{\phi_t L_t }_\Q \, d \left(\frac{1}{Z_t} \right) +  \Braket{d(\phi_t L_t) }_\Q \, d \left(\frac{1}{Z_t}\right), \label{eq:Intro Filt - posterior expectation evolution general}
\end{eqnarray}
where introduced the short-hand notation $\E_\Q\left[ \cdot | \Y_{0:t} \right] = \Braket{\cdot}_\Q$ for the conditional expectation.
We recall from Section~\ref{Sect2} that for both of the signal models, we may write the time evolution of $\phi_t=\phi(X_t)$ as
\begin{eqnarray}
	d \phi_t& = & \mathcal{A}\phi_tdt+dM^{\phi}_t, \nonumber
\end{eqnarray}
where we denote $\mathcal{A}\phi_t=\mathcal{A}\phi(X_t)$.
$M^{\phi}_t$ is a martingale that is independent of the observations under $ \Q $, and thus $\Braket{dM_t^\phi}_\Q = 0$ as well as $\Braket{L_t \, dM_t^\phi}_\Q = 0$.
Therefore, we only retain the $dt$ term under the conditional expectation.
The first term in Eq.~\eqref{eq:Intro Filt - posterior expectation evolution general} can then be computed using Eq.~\eqref{eq:Intro Filt - RadonNikodym evolution}:
\begin{eqnarray}
\Braket{ d(\phi_t L_t) }_\Q & = & \Braket{ (d\phi_t) \, L_t + \phi_t \, (dL_t) + (d\phi_t) \, (dL_t) }_\Q. \nonumber \\
& =  & \Braket{L_t \mathcal{A}\phi_t}_\Q dt + \Braket{\phi_t L_t \tilde{h}(X_t)^{\top}  d\tilde{Y}_t }_\Q, \label{eq:Intro Filt - unnormalized expectation evolution}
\end{eqnarray}
which is the SDE of the \emph{unnormalized} posterior expectation.
Here we further used that $\Braket{ (d\phi_t) \, (dL_t)} = 0 $, because the noise in state and observations are independent.

Note that the evolution equation of the normalization constant $ Z_t $ in Eq.~\eqref{eq:Intro Filt - Kallianpur Striebel},  $ dZ_t = d \Braket{L_t}_\Q $, corresponds to Eq.~\eqref{eq:Intro Filt - unnormalized expectation evolution} with the constant function $ \phi = 1 $.
The differential $ d Z_t^{-1} $ is obtained by consecutive application of It\^{o}'s lemma (Eq.~\ref{eq:Appendix - Ito lemma}).
By plugging Eq.~\eqref{eq:Intro Filt - unnormalized expectation evolution} and $ d Z_t^{-1} $ into Eq.~\eqref{eq:Intro Filt - posterior expectation evolution general} and rewriting everything in terms of expectations under $\P$ using Eq.~\eqref{eq:Intro Filt - Kallianpur Striebel}, one finally obtains the evolution equation for the posterior expectation, the so-called \textbf{Kushner-Stratonovich equation}  (KSE, \citealp[p.~68, Theorem 3.30]{Bain2009}, cf.~\ref{sec:Appendix - KSE} for calculation steps):
\begin{eqnarray}
d\Braket{\phi_t}_\P & = & \Braket{\mathcal{A}\phi_t}_\P \, dt + \cov_\P(\phi_t,\tilde{h}(X_t)^{\top}) (d\tilde{Y}_t -  \Braket{ \tilde{h}(X_t)}_\P \, dt ). \label{eq:Intro Filt - Kushner Stratonovich equation}
\end{eqnarray}
Equivalently, it can straightforwardly be expressed in terms of the original observation process in Eq.~\eqref{eq:Intro Filt - GenerativeModel Observations}:
\begin{eqnarray}
	d\Braket{\phi_t}_\P & = & \Braket{\mathcal{A}\phi_t}_\P \, dt + \cov_\P(\phi_t,h(X_t)^{\top}) \Sigma_y^{-1} (dY_t- \Braket{ h(X_t)}_\P\, dt ).
\end{eqnarray}

In analogy to the calculations in Section~\ref{Sect2}, one may also pass from an evolution equation for the expectations to an adjoint equation for the conditional probability density,
\begin{multline}
	d p (x | \Y_{0:t} ) = \mathcal{A}^{\dag}{p (x | \Y_{0:t} )}\,dt \\
	+ p (x | \Y_{0:t} ) ( h(x)-\Braket{h(X_t)} )^{\top} \Sigma_y^{-1} (dY_t - \Braket{h(X_t)}dt). \label{eq:Intro Filt - Kushner equation}
\end{multline}

Writing Eq.~\eqref{eq:Intro Filt - Kushner Stratonovich equation} and \eqref{eq:Intro Filt - Kushner equation} in terms of the (adjoint of the) infinitesimal generator of the signal process, allows us to use any signal process for which $\mathcal{A}$ is known.
For instance, if the signal process is a Markov chain on a finite set $ S $, the expression $p (x | \Y_{0:t} )$ can be interpreted as the vector of posterior probabilities $\hat{p}(t)$, with entries $\hat{p}_i(t)$ denoting the probability to be in state $i$ given all observations up to time $t$. 
The generator $\mathcal{A}^{\dag}$ is then represented by the matrix $A^{\top}$ that has appeared in the evolution equation for the prior density, Eq.~\eqref{eq:Intro Filt - AdjointEqMC}.
Specifically, $\hat{p}_i(t)$, evolves as
\begin{eqnarray}
	d \hat{p}_i(t) & =& \sum_{j=1}^n A^{\top}_{ij}\hat{p}_j(t)\,dt \nonumber \\
	&&+ \hat{p}_i(t) ( h_i- h \hat{p}(t) )^{\top} \Sigma_y^{-1} (dY_t - h \hat{p}(t)\,dt), \label{eq:Intro Filt - WonhamFilter}
\end{eqnarray}
where $h_i=h(i)\in\mathbb{R}^l$, $i=1,...,n$ and $h$ is an $l\times n$-matrix whose columns are the $h_i$'s.
Eq.~\eqref{eq:Intro Filt - WonhamFilter} is known as the Wonham filter \citep{Wonham1964}, and it is a finite-dimensional SDE that completely solves the filtering problem.

Equation~\eqref{eq:Intro Filt - Kushner equation} is a stochastic integro-differential equation, known as Kushner-Stra\-to\-no\-vich equation (KSE) \citep{Stratonovich1960,Kushner1964}, and its solution is in general infinite-dimensional.
This fundamental problem is easily illustrated by considering the time evolution of the first moment, using $ \phi(x) = x $:
\begin{eqnarray}
	\Braket{X_t} & = & \Braket{f(X_{t})} dt +  \cov_\P( X_t ,h(X_t)^{\top}) \Sigma_y^{-1} (dY_t- \Braket{ h(X_t)}_\P\, dt ). \label{eq:Intro Filt - First moment evolution}
\end{eqnarray}
For non-trivial (i.e.~non-constant) observation functions $ h $, any moment equation will depend on higher-order moments due to the posterior covariance between the observation function $ h $ and the function $ \phi $, which effectively amounts to a closure problem when $f(x)$ is nonlinear.
This is not surprising; even the Fokker-Planck equation \eqref{eq:Appendix - Fokker Planck equation} (the evolution equation for the prior distribution) presents such a closure problem.
In some cases (e.g. when using kernel or Galerkin methods), it is more convenient to use the evolution equation of the \emph{unnormalized} posterior density $ \varrho (X_t | \Y_{0:t}) $:
\begin{eqnarray}
	d \varrho (x | \Y_{0:t} ) & =& \mathcal{A}^{\dag}{ \varrho (x | \Y_{0:t} )}\,dt + \varrho (x | \Y_{0:t} ) h(x)^{\top} \Sigma_y^{-1} dY_t, \label{eq:Intro Filt - Zakai equation}
\end{eqnarray}
which is a linear stochastic partial differential equation (SPDE), the \textbf{Zakai equation} \citep[named after][]{Zakai1969}.

In very rare cases, under specific signal and observation models, the moment equations close, e.g.~in the Kalman-Bucy filter \citep[][see Section \ref{sec:Intro Filt - Kalman Bucy filter} below]{Kalman1961} or the Bene\v{s} filter \citep{BeneS1981}.
Other finite-dimensional filters include the Daum filter \citep{Daum1986} for continuous-time processes and discrete-time measurements.
However, in most cases that occur in practice, the KSE needs to be approximated using a finite-dimensional realization.
For instance, one could use the KSE as a starting point for these approximations, e.g.~Markov-chain approximation methods \citep{Kushner2001} or projection onto a finite-dimensional manifold \citep{Brigo1998,Brigo1999}, which can be shown to be equivalent to assumed density filtering (ADF), or Galerkin-type methods with specific metrics and manifolds \citep[see][]{Armstrong2013}.
Other numerical algorithms associated with overcoming the numerical burden of solving the Kushner or Zakai equation rely on a Fourier approximation of the involved densities, and the fact that convolutions correspond to simple products in Fourier space \citep{Mikulevicius2000,Brunn2006,Jia2010}.

\subsection{A closed-form solution for a linear model: Kalman-Bucy filter}

\label{sec:Intro Filt - Kalman Bucy filter}

In models where the hidden drift function $ f(X) $ and the observation function $ h(X) $ are linear, i.e.~$ f(x) = A x $ and  $ h(x) = B x $, and the initial distribution is Gaussian, there exists a closed-form solution to the filtering problem.
In this case, the KSE (Eq.~\ref{eq:Intro Filt - Kushner Stratonovich equation}) closes with the second posterior moment $ \Sigma_t $, i.e.~the evolution equation for $ \Sigma_t $ becomes independent of the observations process, and the posterior density corresponds to a Gaussian with time-varying mean $ \mu_t $ and variance $ \Sigma_t $.
The dynamics of these parameters are given by the \textbf{Kalman-Bucy filter} (KBF, \citealp{Kalman1961}) and form a set of coupled SDEs:
\begin{eqnarray}
d \mu_t &=& A \mu_t \, dt + \Sigma_t B^{\top} \Sigma_y^{-1} \big( dY_t - B \mu_t \, dt \big),\\
d \Sigma_t & =  & \Big( A \Sigma_t + \Sigma_t A^{\top} + \Sigma_x - \Sigma_t B^{\top} \Sigma_y^{-1} B \Sigma_t \Big)\, dt.
\end{eqnarray}
The posterior variance follows a differential Riccati equation and, due to its independence from observations as well as from the posterior mean, it can be solved offline.


\subsection{Filtering equations for observations corrupted by Poisson noise}
\label{sec:Intro Filt - The filtering problem with point-process observations}
In analogy to the previous section, the formal solution to the filtering problem with observations corrupted by Poisson noise (Eq.~\ref{eq:Intro Filt - Generative Model PP Observations}) can also be derived with the change of probability measure method.
We will very briefly outline the derivation, referring to similarities to continuous-time derivations.\footnote{A very detailed derivation is offered in \citet[p.~170ff.]{bremaud1981point} or, more intuitively, in \citet[SI]{Bobrowski2009} and in \citet[p.~41ff]{Surace2015}.}

The idea is again to make use of the Kallianpur-Striebel formula (Eq.~\ref{eq:Intro Filt - Kallianpur Striebel}) and to rewrite the posterior expectation under the original measure $ \E_\mathbb{P} [\phi_t | \N_{0:t}] $ in terms of an expectation under a reference measure $ \Q $, under which hidden process $ X_t $ and observation process $ N_t $ are decoupled.
Using a Girsanov-type theorem for point processes \citep[see][Chapter VI, Theorems T2 and T3]{bremaud1981point}, the measure is changed to the reference measure $\Q$ under which all point processes have a constant rate $\lambda_0$.
The Radon-Nikodym derivative reads
\begin{eqnarray}
L_t & = & 
\prod_{i = 0}^l  \exp \left( \int_{0}^{t} \log\frac{h_i(X_{s})}{\lambda_0} dN_{i,s} +  \int_{0}^{t} (\lambda_0 - h_i (X_s)  ) \, ds \right) , \label{eq:Intro Filt - RadonNikodym PP}
\end{eqnarray}
which solves the SDE
\begin{eqnarray}
dL_t & = & L_t \cdot \sum_{i = 1}^{l} \left( \frac{h_i(X_t)}{\lambda_0}  -1 \right) (dN_{l,t} - \lambda_0 dt). \label{eq:Intro Filt - RadonNikodym evolution PP}
\end{eqnarray}
We can now repeat the calculations of the previous section.
First, we obtain
\begin{eqnarray}
\Braket{ d(\phi_t L_t) }_\Q & = & \Braket{ (d\phi_t) \, L_t + \phi_t \, (dL_t) + (d\phi_t) \, (dL_t) }_\Q \\
& = & \Braket{ \mathcal{A}\phi_t  \, L_t}_\Q dt \nonumber \\
& & + \Braket{\phi_t  L_t \cdot \sum_{i = 1}^{l} \left( \frac{h_i(X_t)}{\lambda_0}  -1 \right) }_\Q (dN_{i,t} - \lambda_0 dt) . \label{eq:Intro Filt - unnormalized expectation evolution PP}
\end{eqnarray}
Here, we used again that under $ \Q $, differentiation and expectation can be interchanged.

Using $ \phi_t =1 $ gives us the evolution of the time-dependent normalization $ Z_t $ in the Kallianpur-Striebel formula \eqref{eq:Intro Filt - Kallianpur Striebel}.
We can use these, together with It\^{o}'s lemma (Eq.~\ref{eq:Appendix - Ito lemma}) to compute $ d Z_t^{-1} = d \Braket{L_t}_\Q^{-1} $, to obtain a point-process observations analogue to the KSE for the normalized posterior estimate:\footnote{See Appendix \ref{sec:Appendix - The filtering problem with PP observations} for detailed derivation steps.}
\begin{eqnarray}
d\Braket{\phi_t}_\P&=& d \left(Z_t^{-1} \Braket{L_t \phi_t}_\Q \right) \nonumber
\\
& = &\frac{1}{Z_t} \, \Braket{ d( \phi_t L_t ) }_\Q + \Braket{\phi_t L_t }_\Q \, d \left( \frac{1}{Z_t} \right) +  \Braket{d(\phi_t L_t) }_\Q \, d \left( \frac{1}{Z_t} \right) \nonumber
\\
& = & \Braket{\mathcal{A}\phi_t}_\P\, dt + \sum_{i=1}^{l}  \frac{ \cov_\P( \phi_t,h_i(X_t))}{\Braket{h_i(X_t)}_\P}  \left(dN_{d,t} - \Braket{h_i(X_t)}_\P\, dt \right) \nonumber 
\\
& = & \Braket{\mathcal{A}\phi_t}_\P\, dt \nonumber
\\ & & + \cov_\P( \phi_t, h(X_t)^\top) \,  \text{diag}( \Braket{h(X_t)}_\P)^{-1} \left(dN_{t} - \Braket{h(X_t)}_\P\,dt \right),
\label{eq:Intro Filt - Kushner Stratonovich PP}
\end{eqnarray}
where $\text{diag}( x )$ denotes a diagonal matrix with diagonal entries given by the vector $x$.
The adjoint form of this equation, i.e. the evolution equation for the posterior density $  p(x | \N_{0:t}) $, reads:
\begin{eqnarray}
 d p(x | \N_{0:t})  & = & \mathcal{A}^{\dag}{p(x | \N_{0:t})} \, dt + \nonumber \\
 & & p(x | \N_{0:t}) \sum_{i = 1}^l \frac{1}{\Braket{h_i(X_t)}}  \left( h_{i}(x) - \Braket{h_i(X_t)} \right) (d N_{i,t} - \Braket{h_i(X_t)} dt) \nonumber
\\
 & = & \mathcal{A}^{\dag}{ p(x | \N_{0:t})} \, dt + \nonumber \\
 & & p(x | \N_{0:t})    \left( h(x) - \Braket{h(X_t)} \right)^{\top} \text{diag}\left(\Braket{h(X_t)} \right)^{-1} (d N_t - \Braket{h(X_t)} dt).
 \label{eq:Intro Filt - Kushner PP}
\end{eqnarray}
Note the structural similarity to the Kushner equation (Eq.~\ref{eq:Intro Filt - Kushner equation}):
it also relies on a Fokker-Planck term denoting the prediction, and a correction term that is proportional to the posterior density, the `innovation' $ dN_t - \Braket{h(X_t)}dt $, as well as a local correction $  h(x) - \Braket{h(X_t)} $.
The difference is that the observation noise covariance $ \Sigma_y $ in the Kushner equation has been replaced by a diagonal matrix whose components are proportional to the rate function in each observable dimension.
Considering that the observations are Poisson processes, this is not surprising: 
for Poisson processes, the variance is proportional to the instantaneous rate, and thus, analogously, the correction term in this equation has a similar proportionality.

Similarly, we find for the unnormalized posterior density $\varrho(x | \N_{0:t}) $:
\begin{eqnarray}
d \varrho(x | \N_{0:t})  & = & \mathcal{A}^{\dag}{\varrho(x | \N_{0:t})} \, dt + \varrho(x | \N_{0:t}) \frac{1}{\lambda_0}  \left( h(x) - \lambda_0 \right)^T (d N_{t} - \lambda_0 dt). \label{eq:Intro Filt: Zakai PP}
\end{eqnarray}
Analogously to Eqs.~\eqref{eq:Intro Filt - Kushner equation} and \eqref{eq:Intro Filt - Zakai equation}, these equations are obtained by integrating the equation for the unnormalized posterior estimate (Eq.~\ref{eq:Intro Filt - unnormalized expectation evolution PP}) and the normalized posterior estimate (Eq.~\ref{eq:Intro Filt - Kushner Stratonovich PP}) twice.

\subsection{Down to Earth - an example from decision making}
\label{sec:down to earth}

\begin{figure}
	\centering
	\includegraphics[width=1\linewidth]{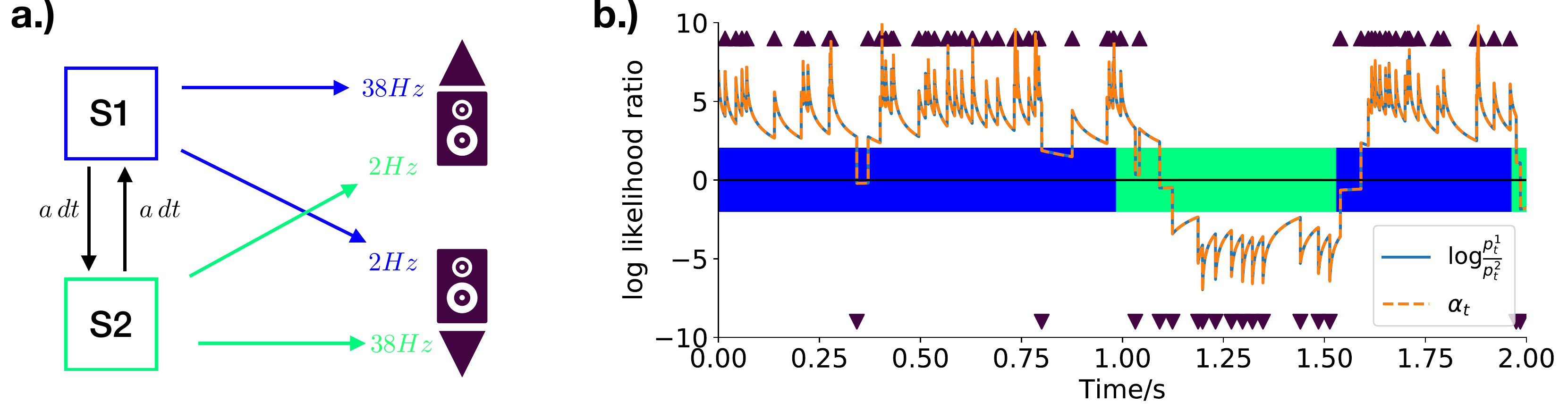}
	\caption{A hidden Markov model with point process observations as an example for a dynamical decision making paradigm. a.) Schematic of the HMM underlying the experiments in \citet{Piet2017}. b.) Logarithm of the likelihood ratio $\alpha_t $, which serves as the decision variable, as a function of time. The sign of the decision variable at the end of the trial denotes the optimal decision.}
	\label{fig:downtoearth}
\end{figure}

In this section, we have seen that the filtering equations can be derived by changing from the original measure to a measure under which the signal and observation processes are independent.
Interestingly, what we have seen here also gives us a recipe of how to treat a filtering problem in general:
all we need is a characterization of the state dynamics in terms of the infinitesimal generator $\mathcal{A}$.
Further, any information about the observations is carried by the Radon-Nikodym derivative.

To illustrate this, let us consider the following example from dynamical decision making.
In order to make a decision that relies on an unobserved, but dynamic, variable, animals have to be able to integrate the (noisy) observations about this variable in a way that discounts older evidence in favor of more recent evidence, depending on the dynamics of the variable.
In other words: they have to do filtering.
In \citet{Piet2017} an experimental paradigm is considered where the hidden variable can be in one of two states, i.e. $X_t \in \{S_1, S_2\} $, which switches states with a hazard rate $ a $ and which influences the click rates $ r_{+} $ and $ r_{-} $ of two speakers.
More precisely, whenever the system is in state $ S_1 $, speaker 1 will have click rate $ r_{+}  $ and speaker 2 will have click rate $ r_{-} $, and vice versa if the system is in state $ S_2 $ (Fig.~\ref{fig:downtoearth}a).
Presented with clicks from these speakers, rats have to make a decision about the state the environment is in at the end of the trial.
The optimal solution to the problem is then the filtering distribution conditioned on the clicks at the end of the trial, and the optimal decision corresponds to the state with the highest posterior probability.
This study has shown that rats are able to perform optimal evidence discounting.\footnote{We are glad they didn't use mice, as these animals, according to the Hitchhiker's guide, are the most intelligent species on planet Earth and as such would surely have outperformed the optimal solution. Rats are close enough, though.}

To come up with the solution for the filtering task, \citet{Piet2017} consider the evolution of the log-likelihood as the decision variable (based on \citealp{Veliz-Cuba2016}) and derive this by taking the continuum-limit of the discrete-time process.
In principle, this approach is perfectly equivalent to a change of measure in the log domain.
Here, we will re-derive their result for the optimal solution of the filtering task by directly applying our `recipe'.

Let us first consider the state process, which is a discrete-state Markov model.
Without observations, the probabilities of the hidden states evolve as $d \tilde{p}_t = \mathcal{A}^\dagger \tilde{p}_t \, dt$, where $\tilde{p}_t^{i} = p(X_t = S_i ) $.
It is easy to check that for this model the adjoint generator matrix is given by
\begin{eqnarray}
	\mathcal{A}^\dagger & = & \begin{pmatrix}
	-a & a \\
	a& -a
	\end{pmatrix},
\end{eqnarray}
where $ a $ denotes the switching rate, i.e. $ a\,dt $ is the probability of switching the state.

The observation model is a two-dimensional point process, and $ N_t^j $ denotes the number of clicks in speaker $ j $ up to time $ t $.
Let $ h_{ji} = h_j(X=S_i) $ be the rate of speaker $ j $ if the hidden variable is in state $ S_i $.
Thus, the evolution of the posterior probability $ p_t $ is given by (cf.~Eq.~\ref{eq:Intro Filt - Kushner PP}):
\begin{eqnarray}
	d p_t^i &=& (\mathcal{A}^\dagger p_t)_i \, dt + p_t^i \sum_j   ( h_{ji}  - \ev{h_j} ) \ev{h_j}^{-1} ( dN_t^j -  \ev{h_j} \, dt), \label{eq:example dpt}
\end{eqnarray}
where $ \ev{h_j} = \sum_i h_{ji} p_t^i $.
Since this particular system is 2-dimensional, i.e.~$ p_{t}^{2} =  1 - p_{t}^{1}$, we can substitute this in the expression for the first component of the posterior and get a one-dimensional equation for $ p_t^1 $.
\begin{eqnarray}
	dp_t^1 & = & a (1-2p_t^1)\, dt  \nonumber \\
	 & & + p_t^1 \sum_{j=1}^2 (h_{j1} - h_{j2})(1-p_t^1)\left(  1/\ev{h_j} \, dN_t^j - dt)  \right), \label{eq:example dpt1}
\end{eqnarray}
where $ \ev{h_j} = h_{j1}p_t^1+h_{j2}(1-p_t^1) $.
At the end of the trial, $ S_1 $ is the optimal choice whenever $ p_t^1 > 1/2 $.

For several reasons, it might be desirable to define the decision variable as the log likelihood ratio of being in state $ S_1 $ as opposed to being in $ S_2 $.
Let $ \alpha_t = \log \frac{p_t^1}{p_t^2} =  \log \frac{p_t^1}{1-p_t^1} $.
In order to derive its evolution equation, we can directly apply It\^{o}'s lemma for point processes to Eq.~\eqref{eq:example dpt1} with $ \phi(x) = \frac{x}{1-x} $ and after straightforward, but tedious, algebra arrive at the desired SDE.
\begin{eqnarray}
	d \alpha_t & = & - 2 a \sinh \alpha_t \, dt \nonumber \\
			&  & + \sum_j \left[ (h_{j2} - h_{j1})\, dt + \log \frac{h_{j1}}{h_{j2}} \, dN_t^j \right].
\end{eqnarray}
Note that if $ h_{11} - h_{12} = h_{22} - h_{21} = r_{+} - r_{-}$ (which is the experimental setting in \citealp{Piet2017}), this equation becomes very simple:
\begin{eqnarray}
d \alpha_t & = & - 2 a \sinh \alpha_t \, dt + \log \frac{r_{+}}{r_{-}} \, (dN_t^1 - dN_t^2), \label{eq:example piet et al}
\end{eqnarray}
and resembles Eq.~9 in \citet{Piet2017}.
Without this symmetry, there is a drift term modulated by the difference in click rates, indicating that the \emph{absence} of clicks is informative for estimating the current state.
In Figure \ref{fig:downtoearth}, we plotted the log likelihood ratio, both computed from the posterior probabilities as in Eq.~\eqref{eq:example dpt}, and directly from running Eq.~\eqref{eq:example piet et al}.
Unsurprisingly, both plots lie exactly on top of each other.

Note that this result was obtained by simply plugging in a model, i.e. the signal and the observation process, into the solution to the filtering equation, and making use of the fact that for a finite state space, the Kushner equation for point processes, Eq.~\eqref{eq:Intro Filt - Kushner PP}, becomes finite-dimensional.
Unlike in \citet{Piet2017} or in \citet{Veliz-Cuba2016}, we did not have to explicitely carry out the continuum limit - in fact, this is implicitely taken care of by using the appropriate Radon-Nikodym derivative for this observation model.
This allows for much more flexibility when the models and/or experimental settings become more complicated, for instance if we want to increase the number of states, modify the state dynamics or modify the properties of the speakers.

A fully annotated code for this example is available in our github repository \citep{Kutschireiter2019a}.

%% file: 04_Approximate-closed-form.tex
\section{Don't panic: Approximate closed-form solutions}

\begin{flushright}
\textit{``It is a mistake to think you can solve any major problems just with potatoes.''}\\
--- Douglas Adams
\end{flushright}

If the signal model is a jump-diffusion, the KSE (Eq.~\ref{eq:Intro Filt - Kushner Stratonovich equation}, Eq.~\ref{eq:Intro Filt - Kushner Stratonovich PP}) is infinite-di\-men\-sion\-al, a fact known as `closure problem'.
In other words, except for some important exceptions for very specific models, such as the KBF or the Bene\v{s} filter \citep{BeneS1981}, solutions to the general filtering problems are not analytically accessible.
Furthermore, unlike for observations following a diffusion process, no closed-form filter for point-process observations is known.
However, there exist important \emph{approximate} closed-form solutions, which address the closure problem by approximating the posterior density in terms of a set number of sufficient statistics.

Here, we will briefly outline some important examples:
first, the Extended Kalman-Bucy Filter and related methods for point-process observations that rely on a series expansion of the functions in the generative model, such that the posterior is approximated by a Gaussian density.
We will further describe Assumed Density Filters, that choose a specific form of the posterior and propagate the KSE according to this approximation.

\subsection{The Extended Kal\-man Bucy Filter and related approaches}
\label{sec:EKBF}

Based on the Kalman-Bucy filter (Section \ref{sec:Intro Filt - Kalman Bucy filter}), the \textbf{extended Kalman-Bucy filter (EKBF)} is an approximation scheme for nonlinear generative models of the form
\begin{eqnarray}
dX_t & = & f(X_t) dt + \Sigma_x^{1/2} dW_t \nonumber 
\\
dY_t & = & h(X_t) dt + \Sigma_y^{1/2} dV_t. \nonumber
\end{eqnarray}
The EKBF approximates the posterior by a Gaussian with mean $ \mu_t $ and variance $ \Sigma_t $, whose dynamics are derived by local linearization (around the mean) of the nonlinearities in the model \citep[p.~338, Example 9.1]{Jazwinski1970}:
\begin{eqnarray}
d \mu_t &=& f(\mu_t)\, dt + \Sigma_t H^\top(\mu_t) \Sigma_y^{-1}  \big( dY_t - h(\mu_t) \, dt \big), \\
d \Sigma_t &=& \Big( F(\mu_t) \Sigma_t + \Sigma_t F(\mu_t)^\top + \Sigma_x - \Sigma_t H^\top(\mu_t) \Sigma_y^{-1} H(\mu_t) \Sigma_t \Big)\, dt,
\end{eqnarray}
where $ F_{ij} = \frac{\partial f_i}{\partial x_j} $ and $ H_{ij} = \frac{\partial h_i}{\partial x_j} $ denote the Jacobian of the hidden drift function and the observation function, respectively.
For models with multimodal posteriors, this approximation often breaks down:
e.g.~if the noise covariance $ \Sigma_y $ is large, the mean of the EKBF tends to `get stuck' in one of the modes.

Similar approximations exist for point-process observations.
One way to achieve this would be to simply construct an EKBF by assuming \emph{Gaussian} noise in the observations, together with the appropriate linearization \citep[see paragraph below Eq.~17 in][]{Eden2007}.
Another way that allows the point-process observations to directly enter the expressions for mean and variance relies on a Taylor expansion in the log domain of the approximated posterior up to second order (see \citealp{Eden2004} for discrete-time and \citealp{EdenBrown2008} for the continuous-time models).
The continuous-time approximate filter for point processes can be seen as the \textbf{point-process analogue of the EKBF} \citep[cf.][extended by nonlinearity in the hidden state process]{EdenBrown2008}:
\begin{eqnarray}
d \mu_t &=& f(\mu_t)\, dt + \Sigma_t \sum_{i=1}^l (\nabla_x \log h_i(x))\rvert_{x = \mu_t}  \big( dN_{i,t} - h_i(\mu_t) \, dt \big), \\
d \Sigma_t &=& \Big( F(\mu_t) \Sigma_t + \Sigma_t F(\mu_t)^\top + \Sigma_x \Big)\, dt \nonumber \\
            & & - \Sigma_t \sum_{i=1}^l \left( \left. \frac{\partial^2 h_i(x)}{\partial x \partial x^\top} \right\rvert_{x = \mu_t} \, dt + S_i\, dN_{i,t}  \right) \Sigma_t,
\end{eqnarray}
with
\begin{eqnarray}
S_i & = & \begin{cases} \left( \Sigma_t - \left( \left. \frac{\partial^2 \log h_i(x)}{\partial x \partial x^\top} \right\rvert_{x = \mu_t} \right)^{-1} \right)^{-1} & \text{if } \left. \frac{\partial^2 \log h_i(x)}{\partial x \partial x^\top} \right\rvert_{x = \mu_t} \neq 0 \\
                0 & \text{otherwise}
        \end{cases}.
\end{eqnarray}

\subsection{Assumed density filtering}

\label{sec:Intro Filt - ADF PP}

The idea of \textbf{assumed density filters} (ADF) is to specify a set of sufficient statistics, which is supposed to approximate the posterior density, derive evolution equations from the KSE, i.e.~from Equations \eqref{eq:Intro Filt - Kushner Stratonovich equation} and \eqref{eq:Intro Filt - Kushner Stratonovich PP}, and approximate expectations within these evolution equations under the initial assumptions.
To be less abstract, consider approximating the posterior density by a Gaussian density.
Then it suffices to derive evolution equations for mean $ \boldsymbol{\mu}_t $ and variance $ \Sigma_t $ of the approximated Gaussian posterior.
In these evolution equations, higher-order moments will enter, which in turn can be expressed in terms of mean and variance for a Gaussian.

As a concrete example, let us consider a Gaussian ADF for point-process observations (the treatment for diffusion-observations is completely analogous).
Consider the SDEs for the first two moments of the posterior (cf.~Eq.~\ref{eq:Intro Filt - Kushner Stratonovich PP}, detailed derivation in Appendix \ref{sec:Appendix - ADF PP}):
\begin{eqnarray}
d \mu_t & = & \ev{ f(X_t) } \, dt + \cov ( X_t , h(X_t)^\top)\, \text{diag} \left(  \expect{h(X_t)} \right)^{-1} \left( dN_{t} - \ev{h_{t}} dt \right), \label{eq:Intro Filt - ADF PP mu}
	\\
d\Sigma_t & = & \left( \cov(f(X_t),X_t^\top) + \cov( X_t,f(X_t)^\top ) + \Sigma_x \right)  dt  \nonumber \\
& & 
+\sum_{i= 1}^l  \frac{1}{\ev{h_i(X_t)}} \left[ \cov(h_i(X_t),X_t X_t^\top) - \cov(h_i(X_t),X_t )  \mu_t^\top - \mu_t  \cov(h_{i}(X_t),X_t^\top )  \right] \nonumber \\
& & \quad \times \left( dN_{i,t} - \ev{h_i(X_t)} dt \right) \nonumber \\
& & - \sum_{i= 1}^l \frac{1}{  \ev{ h_i(X_t) }^2 } \cov( h_i(X_t), X_t ) \cov( h_i(X_t), X_t^\top ) dN_{i,t}. \label{eq:Intro Filt - ADF PP Sigma}
\end{eqnarray}
The effective realization of the ADF will crucially depend on the specifics of the signal model defined by Eq.~\eqref{eq:Intro Filt - JumpDiffSignal} and the observation model \eqref{eq:Intro Filt - Generative Model PP Observations}, respectively.
For example, \citet{Pfister2009} consider an exponential rate function $ h(x) \propto \exp( \beta x )  $, which leads to a variance update term that is independent of the spiking process.
Of particular interest for decoding tasks in neuroscience are ADFs with Gaussian-shaped rate function (e.g.~\citealp{Harel2018}).

However, for some models ADFs cannot be computed in closed form.
Consider for simple example a rate function that is a soft rectification of the hidden process, e.g.~$ h (x) = \log( \exp(x) + 1 ) $, which, when taking expectations with respect to a Gaussian, does not admit a closed-form expression in terms of mean $ \mu_t $ and variance $ \Sigma_t $.

%% file: 05_Particle-filters.tex
\section{Approximations without Infinite Improbability Drive: Particle Methods}
\label{sec:Intro Filt - Particle methods as approximate solutions to the filtering problem}

\begin{flushright}
\textit{``Each particle of the computer, each speck of dust\\held
within itself, faintly and weakly, the pattern of the whole.''}\\
--- Douglas Adams
\end{flushright}

Particle filtering (PF) is a numerical technique to approximate solutions to the filtering problem by a finite number of samples, or `particles', from the posterior.
Thus, they serve as a finite-dimensional approximation of the KSE, overcoming the closure problem.
The true posterior is approximated by the empirical distribution formed by the particle states $ X_{t}^{(i)} $, i.e.~a sum of Dirac-delta functions at the particle positions $\delta( x - X_t^{(i)})$,  and, if it is a weighted PF, weighted by their corresponding importance weights $ w_t^{(i)}  $,  
\begin{eqnarray}
	p( x |\Y_{0:t})  & \approx& \sum_{i=1}^M w_t^{(i)} \delta ( x - X_t^{(i)} ),
\end{eqnarray}
with $\sum_i w_t^{(i)} = 1$ ensuring normalization. Consequently,
\begin{eqnarray}
	\E_\P\left[ \phi (X_t) | \Y_{0:t} \right] \approx \sum_{i=1}^M w_t^{(i)} \phi(X_t^{(i)}).
\end{eqnarray}

The rationale is based on a similar idea as using the Euler-Maruyama scheme to numerically solve the Fokker-Planck equation and its associated equation for the posterior moments.
As a numerical recipe \citep[for instance provided by][for discrete-time models]{Doucet2000,Doucet2009}, it is easily accessible, because in principle no knowledge of the Fokker-Planck equation, nonlinear filtering theory or numerical methods for solving partial differential equations is needed. 

In this section, weighted particle filters will be introduced from a continuous-time perspective based on the change of probability measure formalism \citep[roughly following][Chapt.~9.1, and extending this to point-process observations]{Bain2009}.
From this formalism, we derive dynamics for the weights and link these to the `curse of dimensionality'.
Finally, to give context for readers more familiar with discrete-time particle filtering, the continuous-time perspective will be linked to the `standard' particle filter (PF).

\subsection{Particle filtering in continuous time}
\label{sec:Intro Filt - Particle filtering in continuous time}

Based on sequential importance sampling, both samples (or `particles') $ X_t^{(i)} $ as well as their respective weights are propagated through time.
As we have seen before in Section \ref{sec:Importance sampling}, importance sampling amounts to a change of measure from the original measure $ \P $ to a reference measure $ \Q $, from which sampling is feasible.
Here, the idea is to change to a measure under which the observation processes are independent of the hidden process, effectively enabling us to sample from the hidden process.
In other words, the particle positions evolve as specified by the infintesimal generator $\mathcal{A}$ of the hidden process (e.g.~Eq.~\ref{eq:Intro Filt - JumpDiffSignal} if the hidden process is a jump-diffusion process).

Why this should be the case is rather intuitive when recalling the Kallianpur-Striebel formula \eqref{eq:Intro Filt - Kallianpur Striebel}:
\begin{eqnarray}
\E_\P\left[ \phi_t | \Y_{0:t} \right] & =& \frac{1}{Z_t} \E_\Q [\phi_t\, L_t | \Y_{0:t} ]. \nonumber
\end{eqnarray}
If we want to approximate the left-hand side of this equation with empirical samples, it would require us to have access to samples from the real posterior distribution, which is usually not the case.
However, since under the measure $ \Q $ on the right-hand side the hidden state and observations are decoupled, the estimate is approximated by empirical samples that correspond to realizations of the hidden process:
\begin{eqnarray}
	\frac{1}{Z_t} \E_\Q [\phi_t\, L_t | \Y_{0:t} ] & \approx & \frac{1}{\bar{Z}_t} \sum_{i=1}^{M} \phi(X_t^{(i)}) L_t (X_t^{(i)}).
\end{eqnarray}
$ \bar{Z}_t = \sum_{i=1}^M L_t (X_t^{(i)}) $ is an empirical estimate of the normalization constant.

Thus, we just need to evaluate the Radon Nikodym derivative at the particle states $ X_{t}^{(i)} $, giving us the importance weight $ w_t^{(i)} $ of particle $ i $ at time $ t $.
For observation corrupted by Gaussian noise (cf.~Eq.~\ref{eq:Intro Filt - RadonNikodym}), this reads:
\begin{eqnarray}
	w_t^{(i)} & = & \frac{1}{\bar{Z}_t} L_t (X_{t}^{(i)}) \\
	 & = & \frac{1}{\bar{Z}_t}  \exp \left[ \int_0^t   h(X_{s}^{i})^\top \Sigma_y^{-1} d Y_s - \frac{1}{2} \int_{0}^{t} h(X_{s}^{i})^\top \Sigma_y^{-1} h(X_{s}^{i}) \, ds \right]. \label{eq:Intro Filt - cont time  weights}
\end{eqnarray}
Similiarly, we find for point-process observations (cf.~Eq.~\ref{eq:Intro Filt - RadonNikodym PP})
\begin{eqnarray}
w_t^{(i)} & = &  \frac{1}{\bar{Z}_t} \prod_{j = 1}^l  \exp \left( \int_{0}^{t} \log\frac{h_j(X_{s}^{(i)})}{\lambda_0} dN_{j,s} +  \int_{0}^{t} (\lambda_0 - h_j (X_s^{(i)})  ) \, ds \right). \label{eq:Intro Filt - cont time  weights PP}
\end{eqnarray}

\subsubsection{Weight dynamics in continuous time}

If one is interested how the weight of particle $ i $ changes over time, it is possible to derive an evolution equation for the particle weights. 
Using It\^{o}'s lemma, we find:
\begin{eqnarray}
d w_t^{(i)} & = & d \left( \frac{L_t(X_t^{(i)})}{\bar{Z}_t}  \right) \nonumber \\
&= & \bar{Z}_t^{-1} dL_t^{(i)} + L_t^{(i)}d\bar{Z}_t^{-1} + dL_t^{(i)} \, d\bar{Z}_t^{-1}, \label{eq:Intro Filt - weight dynamics general}
\end{eqnarray}
For continuous-time observations, (cf.~Eq.~\ref{eq:Intro Filt - RadonNikodym evolution}) yields
\begin{eqnarray}
	dL_t^{(i)}  & = & L_t^{(i)} (h(X_t^{(i)})) ^\top \Sigma_y^{-1} dY_t, \\
	d \bar{Z}_t & = & \sum_{i=1}^M dL_t^{(i)} = \bar{Z}_t \, (\bar{h}_t)^\top \Sigma_y^{-1} dY_t, \label{eq:Intro Filt - dZ}
\end{eqnarray}
where $ \bar{h}_t :=  \sum_i w_t^{(i)} h(X_t^{(i)}) = \bar{Z}_t^{-1} \sum_i L_t^{(i)} h(X_t^{(i)})   $ is the weighted estimate of the observation function $ h_t $ (i.e. under the original measure $ \Pmeasure $).
Applying It\^{o}'s lemma on Eq.~\eqref{eq:Intro Filt - dZ} to obtain $ d\bar{Z}_t^{-1} $, we find for the dynamics of the weights
\begin{eqnarray}
	dw_t^{(i)} & = & w_t^{(i)} \left( h(X_t^{(i)}) -\bar{h}_t \right)^\top \Sigma_y^{-1} (dY_t - \bar{h}_t dt) \label{eq:Intro Filt - weight dynamics}
\end{eqnarray}

Similarly, with Eq.~\eqref{eq:Intro Filt - RadonNikodym evolution PP} we find for point-process observations:
\begin{eqnarray}
dL_t^{(i)} & = & L_t^{(i)} \sum_{j=1}^{l} \frac{1}{\lambda_0} \left( h_j (X_t^{(i)}) - \lambda_0 \right) \left(dN_{j,t} - \lambda_0 \, dt  \right), \\
d\bar{Z}_t &=& \bar{Z}_t \sum_{j=1}^{l}   \frac{1}{\lambda_0} (\bar{h}_j - \lambda_0) \left(  dN_{j,t} - \lambda_0 \, dt \right),
\end{eqnarray}
and thus, using It\^{o}'s lemma for point processes to obtain $ d\bar{Z}_t^{-1} $, with Eq.~\eqref{eq:Intro Filt - weight dynamics general}:
\begin{eqnarray}
dw_t^{(i)} & = & w_t^{(i)} \sum_{j=1}^{l}  \frac{1}{\bar{h}_{j,t}} \left( h_j(X_t^{(i)}) - \bar{h}_{j,t} \right) \left(  dN_{j,t} - \bar{h}_{j,t} dt \right)
\\
& = & w_t^{(i)}   \left( h(X_t^{(i)}) - \bar{h}_{t} \right)^\top \text{diag} (\bar{h}_{t})^{-1} \left(  dN_{t} - \bar{h}_{t} dt \right).  \label{eq:Intro Filt - weight dynamics PP}
\end{eqnarray}

Interestingly, there is a striking similarity to the dynamics of the importance weights and the Kushner equation \eqref{eq:Intro Filt - Kushner equation} and the point-process observation equivalent of the Kushner equation (Eq.~\ref{eq:Intro Filt - Kushner PP}), respectively.
The weight dynamics seem to directly correspond to the dynamics of the correction step (with true posterior estimates replaced by their empirical counterparts).
This is rather intuitive:
since we chose a change of measure under which the particles follow the hidden dynamics, serving as the prediction step, the observation dynamics have to be fully accounted for by the weight dynamics, in a way to be consistent with the Kushner equation.


It is important to note that the weight dynamics inevitably lead to a system, in which all but one weight equals zero, the so-called weight degeneracy.
In this degenerate state, the particle system cannot represent the posterior sufficiently, and hence has to be avoided numerically, e.g.~by resampling the particles from the weight distribution and resetting the weights to $1/M$.
The time scale on which this weight degeneracy happens depends on the number of observable dimensions, in other words it is accelerated as the dimensionality of the system is increased \citep{Surace2017}.
This is a form of the so-called 'curse of dimensionality', a common nuisance in weighted particle filters.

\subsubsection{Equivalence between continuous-time particle filtering and bootstrap particle filter}
\label{sec:Intro Filt - Equivalence between continuous-time particle filtering and bootstrap particle filter}

The practitioner who is using PF algorithms in their numerical implementations might usually be more familiar with the discrete-time formulation.
Further, since the con\-ti\-nu\-ous-time formulation of the particle filter based on the measure change formalism seems to be so different from the discrete-time formulation, one might rightfully ask how these two concepts are related and whether they are equivalent.
Indeed, we will now quickly show that the continuous-time PF in Section \ref{sec:Intro Filt - Particle filtering in continuous time} corresponds to the Bootstrap PF in a continuous-time limit.
More precisely, if we apply the Bootstrap PF to a time-discretized version of our hidden state process model and observation model, and then take the continuum limit, we will regain the equations for the weights as in Section \ref{sec:Intro Filt - Particle filtering in continuous time}.

Irrespectively of the generator $ \mathcal{A} $ of the hidden process, it is straightforward to write the hidden process in terms of a transition density, with $ t-dt $ corresponding to the previous time step.
This acts as the proposal density $ \pi (X_t | X_{0:t-dt}^{(i)}, \Y_{0:t}) = p(X_t|X_{t-dt}^{(i)} ) $.
Consider for example a drift-diffusion process (Eq.~\ref{eq:Intro Filt - JumpDiffSignal} with $ J=0 $).
Then the particles are sampled from the time-discretized transition density, $ X_t^{(i)} \sim  p(X_t|X_{t-dt}^{(i)} )  $, which is given by:\footnote{Remark: The so-called Euler-Maruyama scheme for numerical implementation of diffusion processes is based on the very same discretization.}
\begin{eqnarray}
p(X_t|X_{t-dt}^{(i)} )  &= &\mathcal{N}\Big(X_t ; X_{t-d t}^{(i)} + f(X_{t-d t}^{(i)})\, dt, \Sigma_x \, dt \Big). \label{eq:Intro Filt - Equ CT PF - transition}
\end{eqnarray}
For observations corrupted by Gaussian noise, the emission likelihood is given by the emission probability for the \emph{instantaneous increments} $ dY_t $ in Eq.~\eqref{eq:Intro Filt - JumpDiffSignal}, i.e.
\begin{eqnarray}
p (dY_t | X_{t}^{(i)})  & = & \mathcal{N} \Big( dY_t ; h(X_t^{(i)})\,dt, \Sigma_y \, dt \Big),
\end{eqnarray}
such that
\begin{eqnarray}
\tilde{w}_t^{(i)} & = & \tilde{w}_{t-dt}^{(i)} \, \mathcal{N} \Big( dY_t ; h(X_t^{(i)})\,dt, \Sigma_y \, dt \Big) . \label{eq:Intro Filt - Equ CT PF - weights}
\end{eqnarray}

It is evident that the proposal of continuous-time particle filtering and that of the BPF match, and it remains to show that the same holds for the importance weights.
In other words, when taking the continuum limit of Eq.~\eqref{eq:Intro Filt - Equ CT PF - weights}, we should be able to recover Eq.~\eqref{eq:Intro Filt - cont time  weights}.
Keeping only terms up to $ \mathcal{O} (dt) $, we find
\begin{eqnarray}
p (dY_t | X_{t}^{(i)})  & \propto & \exp \left( -\frac{1}{2} (dY_t - h(X_t^{(i)}) dt )^\top (\Sigma_y dt) ^{-1} (d Y_t - h(X_t^{(i)}) dt)\right) \nonumber \\
& \propto &  \exp\left( h(X_t^{(i)})^\top \Sigma_y^{-1} dY_t - \frac{1}{2} h(X_t^{(i)})^\top \Sigma_y^{-1} h(X_t^{(i)}) dt \right),
\end{eqnarray}
where the term $ \propto (d Y_t)^2 $ was absorbed in the normalization because it is independent of the particle positions.
Thus, the continuous-time limit $ dt \to 0 $ of Eq.~\eqref{eq:Intro Filt - bootstrap weights unnormalized} reads
\begin{eqnarray}
\tilde{w}_t^{(i)} & \propto & \prod_{s=0}^{t} \tilde{w}_s^{(i)} = \prod_{s=0}^{t} \exp\Bigg( h(X_s^{(i)}) )^\top \Sigma_y^{-1} dY_s- \frac{1}{2} h(X_s^{(i)})^\top \Sigma_y^{-1} h(X_s^{(i)}) ds \Bigg) \nonumber \\
& \to &  \exp\left( \int_{0}^{t}h(X_s^{(i)})^\top \Sigma_y^{-1} dY_s - \frac{1}{2} \int_{0}^{t}h(X_s^{(i)})^\top \Sigma_y^{-1} h(X_s^{(i)}) ds \right),
\end{eqnarray}
which, up to the normalization constant $ \bar{Z}_t $, is equivalent to Eq.~\eqref{eq:Intro Filt - cont time  weights}.

For point-process observations, the emission likelihood is given by $ p(dN_t|X_{t}) $, which is defined by the Poisson density in Eq.~\eqref{eq:Intro Filt - Generative Model PP Observations}.
Neglecting the term that is independent of the particle positions (which is absorbed in the normalization), it can be rewritten as:
\begin{eqnarray}
	p(dN_t|X_{t}^{(i)}) &=& \prod_j \text{Poisson}(dN_{j,t};h_j(X_t^{(i)})  dt)  \nonumber
	\\
	& = & \prod_j \frac{1}{dN_{j,t}!} \exp\left( - h_{j}(X_t^{(i)}) dt + dN_{j,t}  \log (h_{j}(X_t^{(i)})dt)\right) \nonumber
	\\
	& \propto & \prod_j \exp\left( \log h_{j}(X_t^{(i)}) dN_{j,t} - h_{j}(X_t^{(i)}) dt \right). \label{eq:Intro Filt - PF emission probability PP}
\end{eqnarray}

Again, since $ \tilde{w}_t^{(i)} \propto \prod_s p(dN_s|X_{s}^{(i)}) $, the continuous-time limit of the unnormalized importance weight is
\begin{eqnarray}
	\tilde{w}_t^{(i)} & \to & \prod_j \exp\left( \int_0^t \log h_j(X_s^{(i)}) dN_{j,s} - h_j(X_s^{(i)}) ds \right).
\end{eqnarray}
The explicit dependence of Eq.~\eqref{eq:Intro Filt - cont time  weights PP} on the reference rate $ \lambda_0 $ can be absorbed in the normalization constant, yielding equivalent expressions for the normalized weights $ w_t^{(i)} $.

\subsection{The Feedback Particle filter}

\label{sec:Intro Filt - FBPF}

In contrast to weighted particle filtering approaches, unweighted approaches for particle filtering exist, for example the Ensemble Kalman (Bucy) Filter \citep{Evensen1994,Bergemann2012}, the Feedback Particle Filter (FBPF, \citealp{Yang2013,Yang2014}), the (stochastic) particle flow filter \citep{Daum2010,DeMelo2015} or the point-process analogue to the FBPF \citep{Surace2019}.
Since these methods do not rely on importance weights in the first place, there is no weight degeneracy.
Unweighted particle filters therefore hold the promise of avoiding the curse of dimensionality \citep[see][]{Surace2017}. 

All of these methods have in common that the posterior is approximated by equally weighted particles, i.e.:
\begin{eqnarray}
p(x|\Y_{0:t}) & \approx & \frac{1}{N}\sum_{i=1}^N \delta(x - X_t^{(i)}).
\end{eqnarray}
Consequently, the observations have to enter the particle dynamics in such a way that the particles are moved towards regions of high posterior density.
As an example, we will outline how this is achieved in the Feedback particle filter.

In the FBPF, the observations directly enter the particle dynamics, which evolve according to the It\^{o} SDE:
\begin{eqnarray}
d X_t^{(i)} & = & \left( f(X_t^{(i)},t) + \Omega(X_t^{(i)},t) \right)\,dt + G(X_t^{(i)},t)^{1/2} dB_t^{(i)} \nonumber \\
& &			 + K(X_t^{(i)},t) \Sigma_y^{-1} \left[ dY_t - \frac{1}{2} \left( h(X_t^{(i)}) + \bar{h}_t \right) dt \right], \label{eq:Intro Filt - FBPF}
\end{eqnarray}
where $ B_t^{(i)} $ are uncorrelated vector Brownian motion processes, $ K(X_t^{(i)},t) $ is the gain matrix, and $  \bar{h}_t = \frac{1}{N} \sum_{i=1}^N h( X_t^{(i)} ) $ denotes the particle estimate of the observation function.
The components of the additional vector-valued drift function $ \Omega(X_t^{(i)},t)  $ are given by \citet{Yang2016}
\begin{eqnarray}
\Omega_l(x,t) & = & \frac{1}{2} \sum_{j = 1}^{d} \sum_{k = 1}^{m} K_{jk} (x,t) \frac{\partial K_{lk}}{\partial x_k}(x,t).
\end{eqnarray}

The gain $ K $ is the solution of a boundary value problem that emerges from an optimal control problem \citep[Eqs.~4, 5]{Yang2016}.
It is chosen such that it minimizes the Kullback-Leibler divergence between the particle distribution and the posterior filtering distribution (conditioned on proper initialization), which leads to the following conditions:
\begin{eqnarray}
	\nabla \cdot \left(p(x,t|\Y_{0:t})\nabla\psi_j(x,t) \right)  & = & - \left( h_j - \bar{h}_j \right) p(x,t|\Y_{0:t})
	\\
	\int \psi_j(x,t) p(x,t|\Y_{0:t}) d x & = & 0,
\end{eqnarray}
with
\begin{eqnarray}
	K_{ij} & = & \frac{\partial \psi_j}{\partial x_i}.
\end{eqnarray}

In general, the gain matrix $ K(x,t) $ cannot be solved for in closed form, and in practical implementations relies on a numerical solution of the Euler-Lagrange boundary value problem \citep{Taghvaei2016}.
For instance, one way to approximate the gain $ K(x) $ is using a Galerkin approximation.
In particular, choosing the coordinate functions as basis functions, the so-called constant gain approximation reads \cite[Eq.~20]{Yang2016}:
\begin{eqnarray}
K (x,t) & = & \frac{1}{N} \sum_{i=1}^N X_{t}^{(i)} \left( h(X_{t}^{(i)}) - \bar{h }_t \right)^\top = K(t).
\end{eqnarray}
In this approximation, the gain is constant with respect to the particle positions, i.e.~each particle has the same gain, but still changes as a function of time.
In this approximation, the additional drift function $ \Omega $ in Eq.~\eqref{eq:Intro Filt - FBPF} is zero.

For a linear state space model the FBPF with constant-gain approximation becomes exact and is identical\footnote{up to small numerical difference when computing the gain} to the ensemble Kalman-Bucy filter (EnKBF, \citealp{Bergemann2012,Taghvaei2017}), which can be shown to be asymptotically exact \citep{Kunsch2013}.
More precisely, for a linear model with $f(x) = A x$, $G(x) = \Sigma_x^{1/2}$ and $h(x) = B  x$, the gain $K(x) $ can be solved for in closed form, using the knowledge that the posterior is Gaussian at all times, and is given by $K = \hat{\Sigma}_t B $.
The particles are thus propagated according to (cf.~Eq \ref{eq:Intro Filt - FBPF})
\begin{eqnarray}
dX_t^{i} & = & A X_t^{i} \, dt + \Sigma_x dB_t^{i} + \hat{\Sigma}_t B \Sigma_y^{-1} \left[ dY_t - \frac{1}{2} B \left( X_t^{i} + \hat{\mu}_t \right) \right]
\end{eqnarray}
where $\hat{\mu}_t$ and $\hat{\Sigma}_t$ denote the posterior mean and variance as estimated from the particle positions. 

\subsection{Particle filters in action}

\begin{figure}
	\centering
	\includegraphics[width=\linewidth]{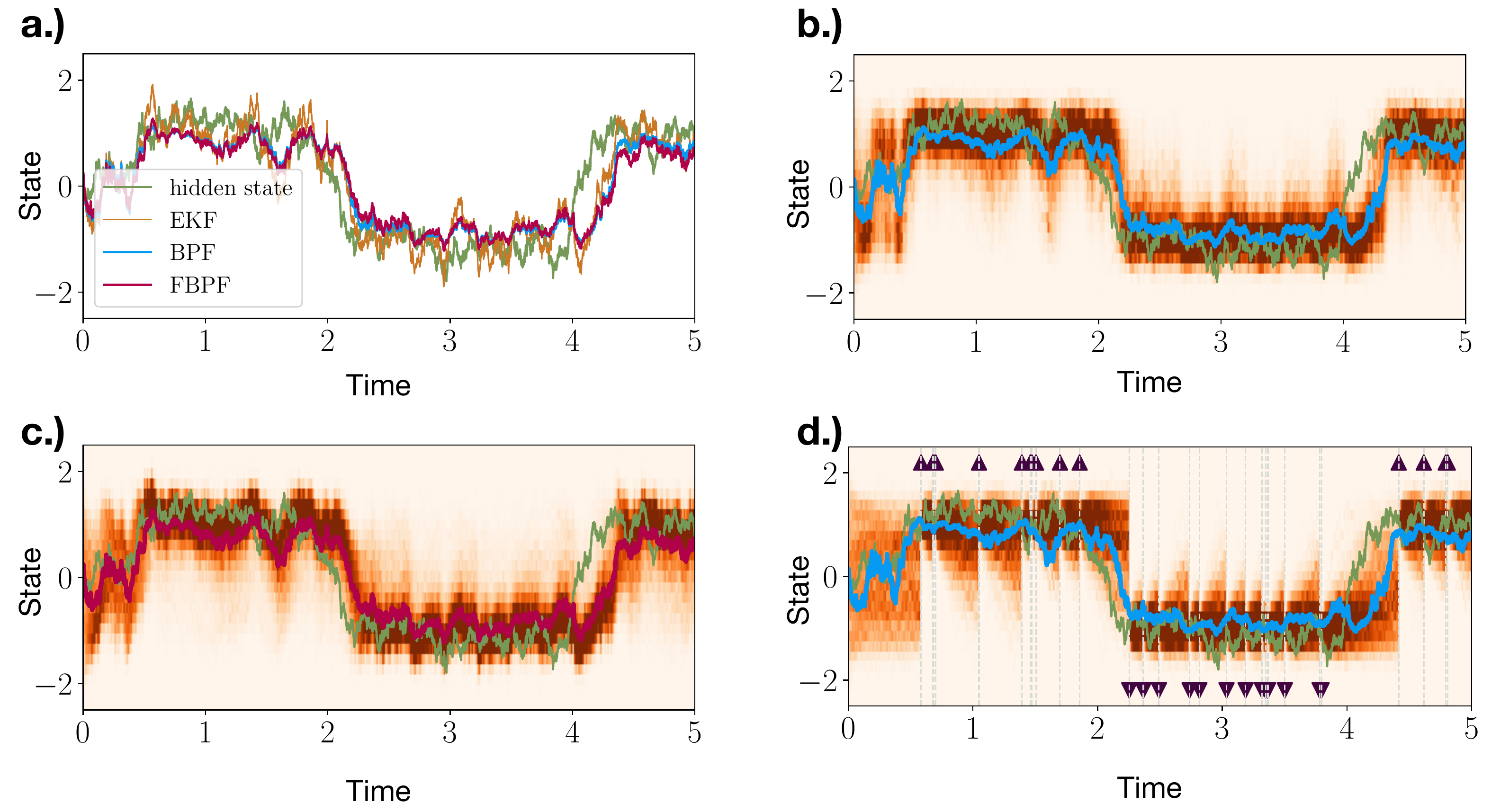}
	\caption{Nonlinear filtering for a hidden state with state dynamics given by $f(x) = -4x(x^2-1)$ and $ \Sigma_x = 2 $. a.) Estimated first posterior moment $ \hat{\mu}_t $ from observations corrupted by Gaussian noise with $ h(x) =1 $ and $ \Sigma_y = 0.1 $. b.) Full posterior from BPF, corresponding to a weighted histogram of the particle positions. c.) Same as b.), but for FBPF. d.) Filtering with the BPF from point-process observations. Here, we consider two sensors with Gaussian-shaped rate functions $ g(x) $ and $ g_0 = 50 $, $ s_0 = 0.05 $ and $ m_0 = \pm1 $. EKF: Kalman-Bucy filter, BPF: Bootstrap particle filter, FBPF: Feedback particle filter with constant-gain approximation.}
	\label{fig:pfinaction}
\end{figure}

Here, we have seen how weighted particle filters can be constructed for nonlinear, continuous time-filtering.
Further, we introduced unweighted particle filters as an alternative, which, for some systems (for instance with a high-dimensional observation model), may be advantageous over standard particle filtering.
However, unweighted particle filters come at the cost of having to compute the gain function, which can be numerically expensive.
Here, we want to demonstrate with an example how these algorithms can be applied as numerical algorithms to a simple nonlinear filtering problem.

Let us consider a hidden process with drift function $f(x) = -4x(x^2-1)$ and diffusion constant $g(x) = \Sigma_x^{-1/2}$.
The corresponding stationary probability density of this nonlinear model is a bimodal distribution, with peaks at $ x = \pm 1 $.
First, we use observations corrupted by Gaussian noise, e.g.~with $ h(x)=x $ (note that the model is still nonlinear due to the nonlinearity in the state transition).
This filtering problem cannot be solved for in closed form and thus we \emph{have to} use a finite-dimensional approximation, such as the particle filter,

With Eqs.~\eqref{eq:Intro Filt - Equ CT PF - transition} and \eqref{eq:Intro Filt - Equ CT PF - weights}, the particle transition and weight dynamics for a standard particle filter (Bootstrap particle filter, BPF) for this model is given by
\begin{eqnarray}
	p(X_t|X_{t-dt}^{(i)} )  &= &\mathcal{N}\Big(X_t ; X_{t-d t}^{(i)}  -4 X_{t-dt}^{(i)} ( (X_{t-dt}^{(i)})^2-1)\, dt, \Sigma_x \, dt \Big), \\
	\tilde{w}_t^{(i)} & = & \tilde{w}_{t-dt}^{(i)} \, \mathcal{N} \Big( dY_t ; X_t^{(i)}\,dt, \Sigma_y \, dt \Big).
\end{eqnarray}
After each iterative step, the weights need to be normalized according to $ w_t^{(i)} = \tilde{w}_t^{(i)} /\sum_j \tilde{w}_t^{(j)} $.
An example tracking simulation is shown in Fig.~\ref{fig:pfinaction}a and b.
For comparison, we also consider an extended Kalman-Bucy filter (see Section \ref{sec:EKBF}) and a feedback particle filter with constant gain approximation (Fig.~\ref{fig:pfinaction}a,c).

Similarly, we might also consider point-process observations with intensity $ g(X_t)$.
For example, let us use a Gaussian-shaped rate function $ g(x) = g_0 \exp(\frac{x-m_o}{2 s_0^2}) $ for two sensors with peaks at $ m_0 = \pm 1 $ and width $ s_0 $.
Figure \ref{fig:pfinaction}d shows that a particle filter is able to track the hidden state reasonably well based only on only the events elicited from these two sensors.
Note also the similarity of this model to the two-state HMM model example we used earlier in Section \ref{sec:down to earth}.

A fully annotated code for this example, which also explains the technical details of this simulation, is available in our github repository \citep{Kutschireiter2019a}.

%% file: 08_Conclusion.tex
\section{The restaurant at the end of the universe: Take-away messages}
\begin{flushright}
\textit{``For where he had expected to find nothing,\\ there was instead a continuous stream of data.''}\\
--- Douglas Adams
\end{flushright}

In this tutorial, we reviewed some of the theory of nonlinear filtering theory.
In fact, we wanted to emphasize that the change of probability measure approach can be used as a universal tool in this context.
Not only does it help to derive the corresponding equations such as the filtering equations, but also leads to a deeper understanding of particle filters.

Let us once more summarize the main take-away messages of this tutorial:
\begin{enumerate}
	\item 
	The change of measure method comes in handy whenever expectations are easier to evaluate under a different measure.
	For example, when computing conditional expectations, it is often easier to compute under a measure in which the random variables are independent. 
	The Radon Nikodym derivative acts as the `conversion factor' between the expectations under the different measures.
	\item 
	The filtering equations can be derived by changing to a reference measure under which signal process and observation process are independent.
	Since this measure change acts on the observation process and leaves the signal process untouched, the filtering equations have the same structure independently of the signal process (which enters in terms of its infinitesimal generator $ \mathcal{A} $).
	Further, all the information about the observations is carried by the Radon-Nikodym derivative.
	\item 
	For a general continuous-time and continous state space nonlinear filtering problem the filtering equations suffer from a closure problem.
	Suitable approximations are based on a finite-dimensional representation of the filtering density, e.g.~in terms of a finite number of statistics (such as the EKF or ADFs) or a finite number of samples (such as PFs).
	\item
	Bootstrap particle filtering can be derived by again changing to a reference measure under which the signal and observation processes are decoupled (which is the very same that was used for deriving the filtering equations), and evaluating the expectations empirically.
	The importance weights correspond to the Radon-Nikodym derivative of this measure change, evaluated at the particle positions.
	Thus, the contribution of the signal process (to the solution of the filtering problem) enters via the positions of the particles (`prediction'), while the contribution of the observations enters via the importance weights (`update').
\end{enumerate}


%% file: 07_Appendix.tex
\newpage
\appendix
\section{Mostly harmless: detailed derivation steps}
\label{chap:Appendix}

\begin{flushright}
\textit{``You ask this of me who have contemplated the very vectors of the atoms of the Big Bang itself? \\Molest me not with this pocket calculator stuff.''}\\
--- Douglas Adams
\end{flushright}

Here, we provide some additional steps that we used in our derivations and left out to keep the main text concise.

\subsection{Evolution equation of Radon-Nikodym derivative $ L_t  $ (Eq.~\ref{eq:Intro Filt - RadonNikodym evolution})}
\label{sec:Appendix - RN evolution}
In Section \ref{sec:Intro Filt - Filtering problem continuous time}, we took the form of the Radon-Nikodym derivative in Eq.~\eqref{eq:Intro Filt - RadonNikodym}, and from this form deduced the evolution equation in Eq.~\eqref{eq:Intro Filt - RadonNikodym evolution}, i.e., Eq.~\eqref{eq:Intro Filt - RadonNikodym} is the solution of Eq.~\eqref{eq:Intro Filt - RadonNikodym evolution}.
Showing this is a nice application of It\^{o}'s lemma and is therefore outlined here in more detail.

Define 
\begin{eqnarray}
	\Lambda_t & : = & \log L_t = \left[ \int_0^t \left(  \Sigma_y^{-1/2} h_s  \right)^\top d \bar{Y}_s - \frac{1}{2} \int_{0}^{t}  \norm{\Sigma_y^{-1/2} h_s}^2\, ds\right]. \\
	d\Lambda_t & = & \left( \Sigma_y^{-1/2} h_t \right)^\top d\bar{Y}_t - \frac{1}{2} \norm{\Sigma_y^{-1/2} h_t}^2\, dt.
\end{eqnarray}
Then, with It\^{o}'s lemma (recall that for diffusion processes this amounts to a Taylor expansion up to second order in the differential):
\begin{eqnarray}
	dL_t &:= & d ( \exp\Lambda_t ) \nonumber
	\\
	& = & \exp \Lambda_t d \Lambda_t + \frac{1}{2} \exp \Lambda_t (d\Lambda_t)^2 \nonumber
	\\
	& = & \exp \Lambda_t d \Lambda_t + \frac{1}{2}  \exp \Lambda_t \norm{\Sigma_y^{-1/2} h_t}^2 dt \nonumber
	\\
	& = &  L_t \left[ \left( \Sigma_y^{-1/2} h_t \right)^\top d\bar{Y}_t - \frac{1}{2} \norm{\Sigma_y^{-1/2} h_t}^2\, dt \right] + \frac{1}{2} L_t \norm{\Sigma_y^{-1/2} h_t}^2 dt  \nonumber
	\\
	& = & L_t  h_t^\top \Sigma_y^{-1/2}  d\bar{Y}_t
	\\
	& = & L_t  h_t^\top \Sigma_y^{-1}  dY_t.
\end{eqnarray}

\subsection{Kushner-Stratonovich Equation (Eq.~\ref{eq:Intro Filt - Kushner Stratonovich equation})}
\label{sec:Appendix - KSE}

The Kushner-Stratonovich equation describes the SDE for the posterior estimate under the original measure $ \P $.
With the Kallianpur-Striebel formula \ref{eq:Intro Filt - Kallianpur Striebel}, we first write the posterior estimate under the reference measure $ \Q $.
This is given by Eq.~\eqref{eq:Intro Filt - posterior expectation evolution general}, revisited here for convenience:
\begin{equation}
d\E_\P \left[ \phi_t | \Y_t \right]  
 =  \frac{1}{Z_t} \, \Braket{ d( \phi_t L_t ) }_\Q + \Braket{\phi_t L_t }_\Q \, d \left( \frac{1}{Z_t} \right) +  \Braket{d(\phi_t L_t) }_\Q \, d \left(\frac{1}{Z_t}\right).  \tag{\ref{eq:Intro Filt - posterior expectation evolution general} revisited}
\end{equation}

Let us evaluate these terms separately.
The first term can be expressed as
\begin{eqnarray}
	\frac{1}{Z_t} \, \Braket{ d( \phi_t L_t ) }_\Qmeasure  & = & \frac{1}{Z_t} \, \Braket{L_t d\phi_t + \phi_t dL_t +  \underset{=0}{\underbrace{d \phi_t dL_t}}}_\Qmeasure \nonumber
	\\
	& = & \frac{1}{Z_t} \left( \Braket{L_t \mathcal{A}\phi_t}_\Qmeasure dt + \Braket{\phi_t L_t (\Sigma_y^{-1/2} h_t)^\top }_\Qmeasure d\bar{Y}_t \right)  \nonumber
	\\ 
	& = & \Braket{\mathcal{A}\phi_t}_\Pmeasure dt + \Braket{\phi_t h_t^\top  }_\Pmeasure \Sigma_y^{-1} d Y_t,
\end{eqnarray}
where we first used the It\^{o} lemma for products, and then made use of $ \Braket{d W_t = 0}_\Qmeasure $.
The term  $ d \phi_t dL_t  $ equals zero because their noise components are independent.
Further, we used the Kallianpur-Striebel formula to rewrite the expressions in terms of expectations under $ \P $.
This equation is basically the unnormalized measure in Eq.~\eqref{eq:Intro Filt - unnormalized expectation evolution} multiplied by $ Z_t^{-1} $, which we could have used directly.

To obtain $ dZ_t^{-1} $, we need to apply It\^{o}'s lemma to the SDE of the normalization constant $ Z_t $ in Eq.~\eqref{eq:Intro Filt - Kallianpur Striebel}.
 $ dZ_t = d \Braket{L_t}_\Q $ is given by to Eq.~\eqref{eq:Intro Filt - unnormalized expectation evolution} with $ \phi = 1 $.
\begin{eqnarray}
dZ_t & = & d \Braket{L_t}_\Qmeasure = \Braket{L_t (\Sigma_y^{-1/2} h_t)^\top d\bar{Y}_t }_\Qmeasure.
\end{eqnarray}
Using It\^{o}'s lemma, we find:
\begin{eqnarray}
d Z_t^{-1} & = & - Z_t^{-2} dZ_t + Z_t^{-3} \left(dZ_t\right)^2 \nonumber
\\
& = & - Z_t^{-2} \Braket{L_t (\Sigma_y^{-1/2} h_t)^\top }_\Qmeasure d\bar{Y}_t  + Z_t^{-3} \Braket{ L_t \norm{\Sigma_y^{-1/2} h_t}}_\Qmeasure^2 dt \nonumber
\\
& = & - Z_t^{-1} \Braket{(\Sigma_y^{-1/2} h_t)^\top }_\Pmeasure d\bar{Y}_t  + Z_t^{-1} \Braket{ \norm{\Sigma_y^{-1/2} h_t}}_\Pmeasure^2 dt \nonumber
\\
& = & - Z_t^{-1} \Braket{h_t}_\Pmeasure^\top \Sigma_y^{-1} (dY - \Braket{h_t}_\Pmeasure dt) \label{eq:Appendix - dZ_t^-1}
\end{eqnarray}
Thus, the second term in Eq.~\eqref{eq:Intro Filt - posterior expectation evolution general} reads:
\begin{eqnarray}
	\Braket{\phi_t L_t }_\Qmeasure \, d \left( \frac{1}{Z_t} \right) & =&  - \Braket{\phi_t}_\Pmeasure \Braket{h_t}_\Pmeasure^\top \Sigma_y^{-1} (dY - \Braket{h_t}_\Pmeasure dt),
\end{eqnarray}

Finally, the third term uses the SDE of the unnormalized posterior expectation in Eq.~\eqref{eq:Intro Filt - unnormalized expectation evolution} and the SDE in Eq.~\eqref{eq:Appendix - dZ_t^-1}, keeping only terms up to $ \mathcal{O}(dt) $.
\begin{eqnarray}
	\Braket{d(\phi_t L_t) }_\Qmeasure \, (d\frac{1}{Z_t}) & = & - \Braket{\phi_t h_t^\top  }_\Pmeasure \Sigma_y^{-1} \Braket{h_t }_\Pmeasure \, dt.
\end{eqnarray}

Adding up and rearranging the terms, we end up with
\begin{eqnarray}
	d\E_\P \left[ \phi_t | \Y_t \right]  & = & \Braket{\mathcal{A}\phi_t}_\Pmeasure dt + \left(  \Braket{\phi_t h_t^\top  }_\Pmeasure - \Braket{\phi_t}_\Pmeasure \Braket{h_t}_\Pmeasure^\top \right) \Sigma_y^{-1} \left( d Y_t -   \Braket{h_t}_\Pmeasure \right) dt \nonumber 
	\\ 
	& = & \Braket{\mathcal{A}\phi_t}_\Pmeasure dt + \text{cov}_\Pmeasure(\phi_t,h_t^\top) \Sigma_y^{-1} \left( d Y_t -   \Braket{h_t}_\Pmeasure dt \right).
\end{eqnarray}

\subsection{Kushner-Stratonovich equation for point-process observations (Eq.~\ref{eq:Intro Filt - Kushner Stratonovich PP})}
\label{sec:Appendix - The filtering problem with PP observations}

The steps are analogous to those taken in the previous section, with a little caveat:
here, SDEs are governed by a point process due to the observation process $ N_t $, so whenever we apply It\^{o}'s lemma, we need to consider an \emph{infinite-dimensional Taylor expansion} in the differential, since $ dN_t^n = dN $.
Also, for simplicity, the following derivation is done for a one-dimensional observation process.
However, it is straightforward to be generalized to $ l $ dimensions by considering that the observations in each dimension are independent conditioned on all observations up to $ t $, which leading to the product in Eq.~\eqref{eq:Intro Filt - RadonNikodym PP} and the sums in Eq.~(\ref{eq:Intro Filt - RadonNikodym evolution PP}ff).

First, using the Kallianpur Striebel formula, we compute the SDE for the normalized posterior expectation by expressing it in terms of the unnormalized posterior SDE:
\begin{eqnarray}
d\Braket{\phi_t}_\P&=& d \left(Z_t^{-1} \Braket{L_t \phi_t}_\Q \right) \nonumber
\\
& = &\frac{1}{Z_t} \, \Braket{ d( \phi_t L_t ) }_\Q + \Braket{\phi_t L_t }_\Q \, d \left(\frac{1}{Z_t}\right) +  \Braket{d(\phi_t L_t) }_\Q \, d \left(   \frac{1}{Z_t} \right),
\end{eqnarray}
where we used It\^{o}'s lemma for products and the fact that under $ \Q $, we can interchange expectation and differentiation.

Again, we compute the terms separately.
Using the evolution equation of the unnormalized measure (Eq.~\ref{eq:Intro Filt - unnormalized expectation evolution PP}, we find:
\begin{eqnarray}
\frac{1}{Z_t} \, \Braket{ d( \phi_t L_t ) }_\Q & = & Z_t^{-1} \Braket{ \mathcal{A}\phi_t  \, L_t}_\Q dt + Z_t^{-1} \Braket{\phi_t  L_t \cdot \left( \frac{h_t}{\lambda_0}  -1 \right) }_\Q (dN_{t} - \lambda_0 dt) \nonumber 
\\
& = & \Braket{ \mathcal{A}\phi_t }_\P dt + \frac{1}{\lambda_0} \left( \Braket{\phi_t L_t}_\P - \Braket{\phi_t}_\P \lambda_0 \right)(dN_{t} - \lambda_0 dt).
\end{eqnarray}

For the second term, we again write down the SDE for the normalization constant $ Z_t $ and its inverse.
From Eq.~\eqref{eq:Intro Filt - unnormalized expectation evolution PP} with $ \phi_t = 1 $ we find:
\begin{eqnarray}
dZ_t = d \Braket{L_t}_\Q & = & \Braket{L_t \left( \frac{h_t}{\lambda_0} - 1 \right)}_\Q (dN_t - \lambda_0 dt) \nonumber \\
& = & \left( \frac{\Braket{h_t}_\P}{\lambda_0} - 1 \right) (dN_t - \lambda_0 dt).
\end{eqnarray}
The SDE for its inverse is obtained by using It\^{o}'s lemma for point processes (Eq.~\ref{eq:Appendix - Ito lemma}):
\begin{eqnarray}
d Z_t^{-1} & = & - Z_t^{-1} ( \lambda_0 - \Braket{h_t}_\P )\,dt + \left[ (Z_t + Z_t \left( \frac{\Braket{h_t}_\P}{\lambda_0} -1  \right)^{-1} - Z_t^{-1} \right] \, dN_t \nonumber \\
& = & Z_t^{-1} \left( \lambda_0 - \Braket{h_t}_\P  \right) \frac{1}{\Braket{h_t}_\P}  \left( dN_t - \Braket{h_t}_\P dt \right). \label{eq:Appendix - dZ_t^-1 PP}
\end{eqnarray}
Now we can write
\begin{eqnarray}
\Braket{\phi_t L_t }_\Q \, d\frac{1}{Z_t} & = & Z_t^{-1} \Braket{\phi_t L_t }_\Q \left( \lambda_0 - \Braket{h_t}_\P  \right) \frac{1}{\Braket{h_t}_\P}  \left( dN_t - \Braket{h_t}_\P dt \right) \nonumber
\\
& = & \Braket{\phi_t }_\P \left( \lambda_0 - \Braket{h_t}_\P  \right) \frac{1}{\Braket{h_t}_\P}  \left( dN_t - \Braket{h_t}_\P dt \right).
\end{eqnarray}

Finally, for the last term, we only keep terms of $ \mathcal{O} (dN) $:
\begin{eqnarray}
\Braket{d(\phi_t L_t) }_\Q \, (d\frac{1}{Z_t}) & = & \frac{1}{\lambda_0} \left( \Braket{\phi_t L_t}_\P - \Braket{\phi_t}_\P \lambda_0 \right) \left( \lambda_0 - \Braket{h_t}_\P  \right) \frac{1}{\Braket{h_t}_\P} \, dN_t.
\end{eqnarray}

Adding up and rearranging the terms, we end up with
\begin{eqnarray}
	d\Braket{\phi_t}_\P&=&  \Braket{ \mathcal{A}\phi_t }_\P dt - \left( \Braket{\phi_t h_t }_\P + \Braket{\phi_t}_\P \right) \, dt +  \left( \Braket{\phi_t h_t }_\P + \Braket{\phi_t}_\P \right) \, dN_t \nonumber 
	\\
	& = & \Braket{ \mathcal{A}\phi_t }_\P dt  + \frac{1}{\Braket{h_t}_\P} \cov_\P (\phi_t,h_t) ( dN_t - \Braket{h_t}_\P\, dt ).
\end{eqnarray}
A generalization to multivariate $ N_t $ and $ h_t $, respectively, is done by treating all observable dimensions separately and summing up.

\subsection{ADF for point-process observations (Eq.~\ref{eq:Intro Filt - ADF PP mu} and \ref{eq:Intro Filt - ADF PP Sigma})}
\label{sec:Appendix - ADF PP}

For the ADF in subsection \ref{sec:Intro Filt - ADF PP}, we need the SDEs for the first two posterior moments.
These can be obtained with the KSE for point processes (Eq.~\ref{eq:Intro Filt - Kushner Stratonovich PP}).
For the mean in Eq.~\eqref{eq:Intro Filt - ADF PP mu}, we use $ \phi(x) = x $:
\begin{eqnarray}
	d \mu_t & = & \Braket{ \mathcal{A} X_t} \, dt + \sum_{i=1}^{l}  \frac{ \cov (h_{i,t} , X_{t})}{\Braket{h_{i,t}} }  \left(dN_{i,t} - \Braket{h_{i,t}} \,dt \right) \nonumber 
	\\
	& = & \Braket{ f_t } \, dt + \sum_{i=1}^{l}  \frac{ \cov (h_{i,t} , X_{t})}{\Braket{h_{i,t}} }  \left(dN_{i,t} - \Braket{h_{i,t}} \,dt \right) \nonumber 
	\\
	& = & \Braket{ f_t } \, dt +   \cov (X_t, h_t^\top ) \text{diag} (\Braket{h_{i,t}})^{-1}  \left(dN_{t} - \Braket{h_{t}} \,dt \right)
\end{eqnarray}

To compute the SDE for the posterior variance, we use It\^{o}'s lemma:
\begin{eqnarray}
	d \Sigma_t = d ( \Braket{X_t X_{t}^\top} - \mu_t \mu_t^\top ) = d \Braket{X_t X_{t}^\top} -\mu_t d\mu_t^\top -  (\mu_t d\mu_t^\top)^\top - (d\mu_t) (d\mu_t)^\top,
\end{eqnarray}
with
\begin{eqnarray}
	d (\Braket{X_t X_{t}^\top} & = & \Braket{f_t X_{t}^\top + X_{t} f_t^\top}dt + \Sigma_x\,dt  \nonumber \\
	& & + \sum_{i=1}^{l}  \frac{ \cov (h_{i,t} , X_{t} X_{t}^\top ) }{\Braket{h_{i,t}} }  \left(dN_{i,t} - \Braket{h_{i,t}} \,dt \right), 
\end{eqnarray}
\begin{eqnarray}
\mu_t d\mu_t^\top & = & \mu_t \Braket{ f_t }^\top \, dt + \sum_{i=1}^{l}  \frac{ \mu_t  \cov (h_{i,t} , X_{t}^\top )}{\Braket{h_{i,t}} }  \left(dN_{i,t} - \Braket{h_{i,t}} \,dt \right),
\end{eqnarray}
\begin{eqnarray}
(\mu_t d\mu_t^\top)^\top & = & \Braket{ f_t } \mu_t^\top \, dt + \sum_{i=1}^{l}  \frac{ \cov (h_{i,t} , X_{t} ) \mu_t^\top  }{\Braket{h_{i,t}} }  \left(dN_{i,t} - \Braket{h_{i,t}} \,dt \right),
\end{eqnarray}
\begin{eqnarray}
(d\mu_t) (d\mu_t)^\top & = & \sum_{i=1}^{l}  \frac{1 }{\Braket{h_{i,t}}^2 } \cov (h_{i,t} , X_{t}) \cov (h_{i,t} , X_{t})^\top dN_{i,t}.
\end{eqnarray}
Adding these up gives us Eq.~\eqref{eq:Intro Filt - ADF PP Sigma}.




%
%